\documentclass[twocolumns]{aa}  

\usepackage{graphicx}
\usepackage{txfonts}
\usepackage{comment}
\usepackage{color}
\usepackage{xcolor}
\usepackage{url}
\usepackage{hyperref}
\usepackage{multirow}
\usepackage{float}
\usepackage{orcidlink}
\usepackage{soul} 

\definecolor{myblue}{HTML}{6a359c} 
\definecolor{myviolet}{HTML}{756BB1}
\definecolor{mygreen}{HTML}{7FCDBB}
\definecolor{myyellow}{HTML}{FFFDD1}

\hypersetup{
	colorlinks,
	linkcolor={myblue},
	citecolor={myblue},
	urlcolor={myblue}}

\newcommand{\pairs}{{\small PAIRS }}
\newcommand{\abin}{a_{\rm bin}}
\newcommand{\ebin}{e_{\rm bin}}
\newcommand{\Rtrunc}{R_{\rm trunc}}
\newcommand{\Regg}{R_{\rm Egg}}

\newcommand{\ME}{\rm{M}$_{\oplus}$ \,} 
 
\newcommand{\Mearth}{\text{M}_{\oplus}} 
\newcommand{\Msun}{\text{M}_{\odot} }

\newcommand{\Mp}{\rm M_{\text{P}}}

\begin{document} 
 \title{The PAIRS project: a global formation model for planets in binaries}
   \subtitle{I. Effect of disc truncation on the growth of S-type planets.}
 \titlerunning{PAIRS-I}
\authorrunning{Venturini et al.}

\author{Julia Venturini\orcidlink{0000-0001-9527-2903}\inst{1},
Arianna Nigioni\inst{1},
Maria Paula Ronco\orcidlink{0000-0003-1385-0373}\inst{2}, 
Natacha Jungo\inst{3}, Alexandre Emsenhuber\inst{3}}

\offprints{J. Venturini}
\institute{Department of Astronomy, University of Geneva, Chemin Pegasi 51, 1290 Versoix, Switzerland, 
\and
Instituto de Astrof\'{\i}sica de La Plata, CCT La Plata-CONICET-UNLP, Paseo del Bosque S/N (1900), La Plata, Argentina.
\and
{Department of Space Research \& Planetary Sciences, University of Bern, Gesellschaftsstrasse 6, CH-3012 Bern, Switzerland}
\\
\email{julia.venturini@unige.ch}}

\date{}

\abstract{Binary stars are as common as single stars. The number of detected planets orbiting binaries is rapidly increasing thanks to the synergy between transit surveys, \textit{Gaia} and high-resolution direct imaging campaigns. However, global planet formation models around binary stars are still underdeveloped, which limits the theoretical understanding of planets orbiting binary star systems. Hereby we introduce the {\small PAIRS} project, which aims at building a global planet formation model for planets in binaries, and to produce planet populations synthesis to statistically compare theory and observations. In this first paper, we present the adaptation of the circumstellar disc to simulate the formation of S-type planets. The presence of a secondary star tidally truncates and heats the outer part of the circumprimary disc (and vice-versa for the circumsecondary disc), limiting the material to form planets. We implement and quantify this effect for a range of binary parameters by adapting the \textit{Bern Model} of planet formation in its pebble-based form and for in-situ planet growth. We find that the disc truncation has a strong impact on reducing the pebble supply for core growth, steadily suppressing planet formation for binary separations below 160 au, when considering all the formed planets more massive than Mars. We find as well that S-type planets tend to form close to the central star with respect to the binary separation and disc truncation radius. 
Our newly developed model will be the basis of future S-type planet population synthesis studies.}

\maketitle
%
\section{Introduction}
Binary stars are very common, half of the Sun-like stars in our galaxy belong to a binary system \citep{Raghavan10}. The detection of planets orbiting binary stars is rapidly increasing, as the sensitivity and resolution of detectors improves. In the last few years, hundreds of planets first detected by transit surveys like \textit{Kepler} and TESS have been identified as belonging to a binary star system thanks to astrometric follow-up by \textit{Gaia} \citep[e.g.][]{Mugrauer2021, Mugrauer2022, Behmard2022, Mugrauer2023} or high-resolution direct imaging \citep[e.g.][]{Lester2021,Sullivan23, Schlagenhauf24}. These systems contain S-type planets, i.e, planets orbiting one of the stars in the pair. The other type of planets in binaries that exists are the circumbinary or P-type.\footnote{“S-type” stems from “satellite”, because the planet is the satellite of one of the stars in the pair; whereas ``P-type" stands from ``planet" since the planet has a ``planetary" orbit around the two stars.} These happen around very close binaries, and in this case the planet orbits both stars. P-type planets are very difficult to detect, we currently know about 30 of them  \citep[e.g.][]{Doyle2011,Triaud2022,Martin2018,Kostov2020,Kostov2021,Socia2020,Standing2023}. On the contrary, today about 500 S-type planets are either confirmed or validated, and another 400 S-type planet candidates from TESS are awaiting confirmation \citep{Mugrauer2021, Mugrauer2022, Mugrauer2023, Lester2021, Hirsch21, Lester2022, Fontanive21}.

With already several hundreds of S-type planets identified, some statistics from observations have been inferred. 
Regarding occurrence rates, by combining direct imaging and radial velocities measurements, \citet{Hirsch21} found that S-type giant planets with semi-major axis between 0.1 and 20 au exist in 20\% of binaries with separations larger than 100 au (a similar occurrence to single stars), but in 4\% of binaries with separations smaller than 100 au, hinting a detrimental effect on giant planet formation for close binaries. By analysing \textit{Gaia} Data Realese 2, \citet{Fontanive21} found that the distribution of planet mass and distance to the host star is statistically different between binaries and single stars as long as the two stars are closer than 1000 au.
In addition, \citet{Lester21} searched for companions to TESS systems known to host planets by direct imaging. They found that contrary to field binaries which have their peak of binary separation at 50 au, for planet-hosting binaries the most frequent binary separation is 100 au. They conclude that this indicates a detrimental effect of planet formation for binaries with separations closer than 100 au. They also found that planets in binaries span a large range of sizes, similar to exoplanets orbiting single stars. More recently, \citet{Thebault_catlog-2025} presented a new Catalog of planets in binaries. Among different properties, they also analysed the distribution of binary separations for systems hosting planets, finding a peak at 500 au. The study also highlights, from recent surveys, that about 20\% of the known exoplanets are in reality S-type planets.
Finally, \citet{Sullivan23, Sullivan24} analysed for the first time the properties of small S-type planet candidates from the \textit{Kepler} survey, in systems where the secondary star was found by direct imaging follow-up. Contrary to single-star planets, which exhibit a \textit{radius valley} separating the population of super-Earths from the population of mini-Neptunes, in the case of S-type planets the mini-Neptunes seem to be suppressed for binary separations below 100 au \citep{Sullivan24}. 

In parallel, ALMA observations of discs around binaries in star forming regions are starting to reveal the detrimental effect that a stellar companion has on the dust mass, the extension of the dust disc and the lifetime of S-type protoplanetary discs \citep{Kutra25, Zurlo23, Zurlo21, Zurlo20, Barenfeld19}.

The rapidly growing population of S-type planets and discs, and the first statistics of S-type planets call for the urgent development of planet formation models targeting the specific environment where S-type planets form. Indeed, such environment can be very different from the single-star case. Namely, the presence of the stellar companion tidally truncates and heats the protoplanetary disc \citep{Papaloizou1977, Arty94, Alexander11}, reducing the reservoir of material to form planets. In addition, the gravitational perturbation from the companion excites the orbits of the growing planets and planetesimals, in principle precluding planetary accretion \citep{Chambers02}. The closer the two stars, the more dramatic the effects \citep[e.g][]{Marzari19}.

The theoretical study of planet formation in binaries is still very limited to a few specific processes, the most explored ones are: i) disc evolution \citep{Alexander11, Rosotti18, Ronco21, Zagaria2021}, ii) planetesimal accretion \citep{Marzari00, Thebault06, Silsbee15b, Silsbee15, Rafikov15a, Rafikov15b}, iii) dynamics after the dissipation of the gaseous disc \citep{Thebault04, Quintana06, Haghi06, Quintana07,Giuppone2011}, and iv) the study of disc structure and planet migration via hydrodynamical simulations \citep{Nelson03, Marzari12, KleyNelson08, Kley14, Kley19, Jordan21}. 
However, some relevant physical processes operating in planet formation, like pebble accretion \citep{OrmelK10, Lambrechts12, LJ14}, have never been included in the formation modelling of S-type planets so far. Pebble accretion has proven to be relevant to account for the short formation timescales of giant planets \citep[e.g.][]{Drazkowska2021}, as well as to explain certain exoplanets' properties \citep[e.g. the dependence of the radius valley with the stellar mass,][]{Venturini24}.

While having reliable individual physical models is essential, it is not enough to compare theory with observations of exoplanets. Indeed, the individual physical processes operating during planet formation, such as core growth, gas accretion, disc evolution and planet migration, happen all at similar timescales and feedback on each other. Because of this, they must be included into a single model if one wants to compare the simulations' output with observations. This is the principle of the so-called \textit{global} or end-to-end \textit{planet formation models} \citep[e.g.][]{Mordasini15}. This principle is also true for planet formation around binary stars. 

Global planet formation models are a collection of 1D models and prescriptions that encapsulate the main processes acting during planet formation. The low dimensionality and simple nature of the individual models is essential to reduce computational time. When these global models are run thousands of times, varying the initial conditions stemming from observations, the output is the so-called `planet population synthesis' \citep{IdaLin04, Mordasini09, Benz14, Fortier13,Ronco17,Emsenhuber21}. Global formation models and population synthesis for S-type planets are at the moment non-existent.
Without global formation models and population synthesis, a quantitative comparison between observations and theory for planets around binaries is not feasible.

Hereby, we introduce the {\small PAIRS} project. {\small PAIRS} stands for "Planet formation Around bInary Stars", and its primary goal is to develop the first global planet formation model for S-type planets, and to compute S-type planet population synthesis. 
For this, we adapt the classical \textit{Bern Model} of planet formation and evolution \citep{Alibert05, Mord09, Fortier13,Emsenhuber21}, to simulate the formation environment of S-type planets. In this first paper, we introduce the physical effects of disc truncation, tidal heating and irradiation from the companion into the evolution of the protoplanetary disc. 
The accompanying paper II (Nigioni et al., submitted) presents the adaptation of the N-body integrator to include the gravitational interaction between all the growing embryos and the stellar companion. 

The outline of this paper is as follows. In Sect. \ref{sec_methods}, we briefly summarise the employed version of the Bern Model and we describe the disc's model modifications for S-type discs. In Sect. \ref{sec_results} we present results of the disc evolution and planetary growth by pebble accretion for a couple of illustrative cases, together with a parameter study. We discuss the limitations of the model in Sect. \ref{sec_discussion} and we summarise our findings in Sect. \ref{sec_conc}.

\section{Methods} \label{sec_methods}
We adapt the Bern Model of planet formation and evolution in its version of \citet[][]{Emsenhuber21}, to model the formation of S-type planets. The Bern Model computes the growth of a planet from a moon-mass embryo by solid and gas accretion, while taking into account the evolution of the disc by viscous accretion and photoevaporation, as well as the evolution of the central star. Once the disc dissipates, the gravitational interactions are computed for 20 Myr, after which the evolution by atmospheric cooling and photoevaporation is computed until 10 Gyr.  
In this work we study planet formation around the primary star. Paper II analyses a set of simulations around the secondary, and we will address the analysis of a population around the secondary in a future work.

In this section, we first dive into the disc model adaptation to simulate the environment where S-type planets form. At the end of this section and in
Appendix \ref{App_extra-model}, we briefly summarise the main physical assumptions of the standard Bern Model that remain unchanged compared to the single-star case.

\subsection{Simple model of disc truncation}\label{sec_disc_truncation}
The main physical effect that a stellar companion produces in the disc of the other star in the binary, is the outer tidal truncation stemming from the gravitational interaction with the companion \citep{Papaloizou1977, Arty94}. %
This tidal truncation of the protoplanetary disc has been studied analytically by \citet{Papaloizou1977} and \citet{Arty94}. 

Numerically, the truncation can be modelled by adding a torque term to the viscous disc evolution equation \citep{Alexander11, Ronco21}. However, numerical instabilities can emerge in such approach when the dust evolution is included. A simplified and equivalent way of modelling the truncation was proposed by \citet{Rosotti18} and \citet{Zagaria2021}. This consists in defining from the beginning of the simulation the outer truncation radius and imposing a zero-flux boundary condition at this location both for the gas and the dust disc. 
We adopt the same approach as \citet{Zagaria2021} and we additionally link the truncation radius to the binary properties by employing the recipe outlined in \citet{Manara19}, where they fit the truncation radius of the S-type discs to the analytical results found by \citet{Arty94}.
Thus, the truncation radius of the circumprimary disc is defined as: 
\begin{equation}\label{eq:Rtrunc}
    R_{\rm trunc} (M_1, M_2, \abin, \ebin)= \Regg \times (b e_{\rm bin}^c +h \mu^k),
\end{equation}
with the Eggleton Radius \citep{Eggleton1983} given by
\begin{equation}\label{eq:Rtrunc2}
    \Regg = \frac{0.49 \, q^{-2/3}}{0.6 \, q^{-2/3} + \text{ln}(1+q^{-1/3})} \quad \abin. 
\end{equation}
$\abin$ is the binary semi-major axis, $\ebin$ the binary eccentricity, and $q=M_2/M_1$ the binary mass ratio, with $M_1$ being the mass of the primary and $M_2$ the mass of the secondary ($M_2<M_1$). In addition, $\mu = M_2/(M_1 + M_2)$ and $b,c,h,k$ are fitting parameters. For the truncation radius of the circumsecondary disc, the formula is the same except that $q$ must be replaced by its inverse, $q'= M_1/M_2$.
The parameters $h$ and $k$ adopt the values $h=0.88$ and $k=0.01$, found by fitting the data from \citet{Papaloizou1977}.
The values of $b$ and $c$ are shown in Table 1, which corresponds to Table C.1 from \citet{Manara19}, and they are given for different values of $\mu$ and Reynolds number $\mathcal{R}$, both for circumprimary and circumsecondary discs. To determine the values of $b$ and $c$ for arbitrary $\mathcal{R}$ and $\mu$, we first compute the Reynolds number as
\begin{equation}\label{eq:Rtrunc3}
\mathcal{R} = \frac{1}{\alpha} \left( \frac{H}{r} \right)^{-2}
\end{equation}
where $\alpha$ is the disk viscosity parameter and $\frac{H}{r}$ is the disc aspect ratio. We fix $\frac{H}{r}=0.0983$, based on the initial radial profiles of a synthetic population of single-star protoplanetary discs. Specifically, we simulate 1000 single-star discs using the \textit{Bern Model} for two values of the viscosity parameter, $\alpha = 10^{-3}$ and $\alpha =10^{-4}$. For each system, we compute the mean value of $\frac{H}{r}$ at orbital distances beyond 10 au, where the aspect ratio radial profiles exhibit reduced scatter. By fitting a Gaussian distribution to the resulting sample of mean values, we find that the central values for the two $\alpha$ disc-populations agree up to the fourth decimal place. We therefore adopt this common value as our reference aspect ratio in the truncation model. We then compute the value of $\log_{10}(\mathcal{R})$, constraining it within the range [4,6] to match the tabulated simulations in \citet{Manara19}, and we perform a two-step interpolation:
\begin{enumerate}
\item For each tabulated value of 
$\mu$, we fit second-degree polynomials in $\log_{10}(\mathcal{R})$ to the corresponding $b$ and $c$ coefficients from \citet{Manara19}.
\item We evaluate these polynomials at the actual Reynolds number to obtain intermediate values of $b$ and $c$ for each $\mu$, and then fit fourth-degree polynomials in $\mu$ to obtain the final values at the desired mass ratio.
\end{enumerate}

\begin{figure}[]
    \centering
    \includegraphics[width=\linewidth]{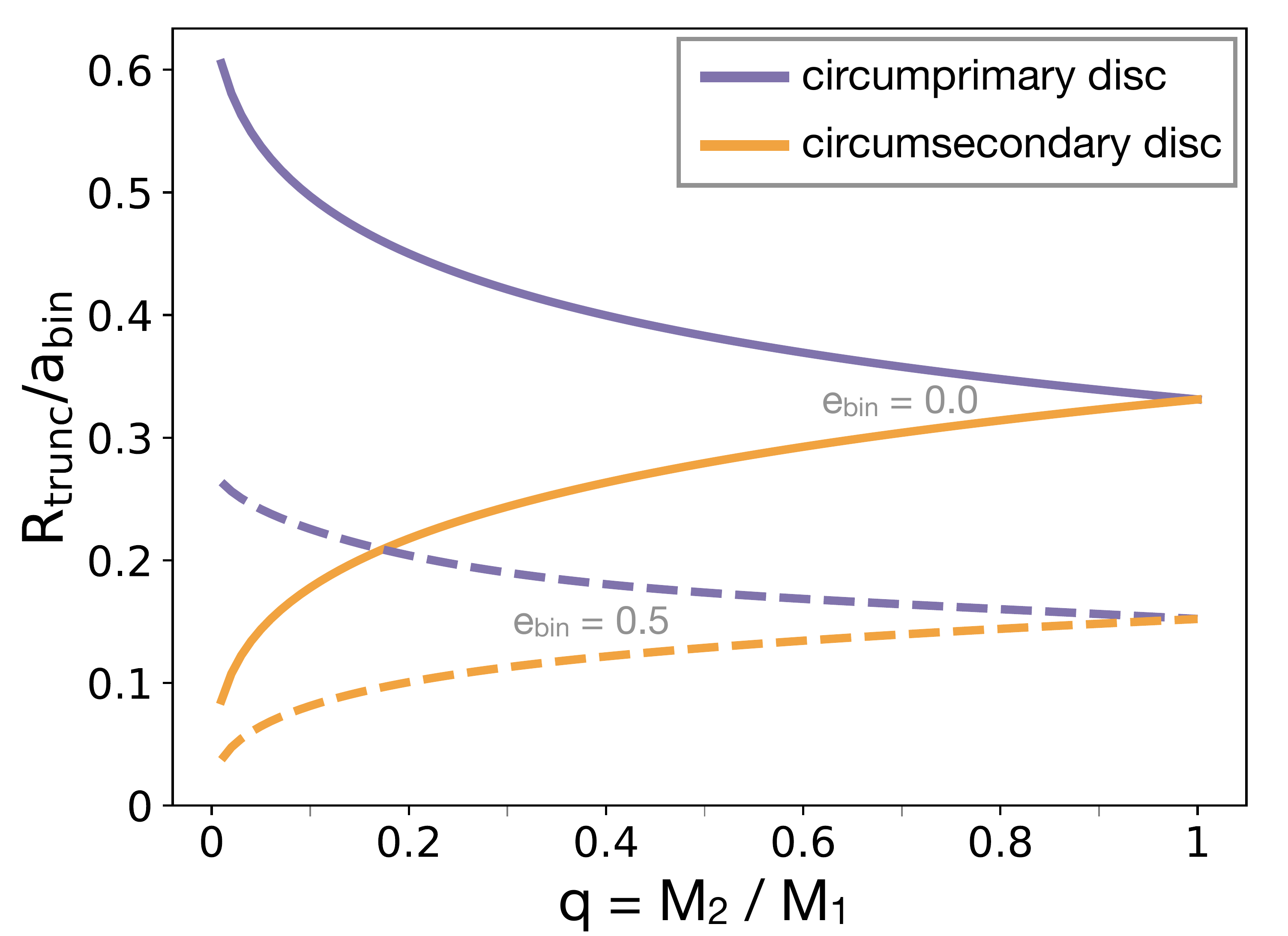}
    \caption{Disc truncation radius in units of the binary semi-major axis as a function of the binary mass ratio ($q$), both for the circumprimary (violet lines) and circumsecondary discs (orange lines). The solid lines corresponds to a binary with $\ebin=0$, while the dashed lines indicate the case where $\ebin=$ 0.5.}
    \label{fig_truncation}
\end{figure}

Figure \ref{fig_truncation} shows the truncation radius computed following the outlined procedure. The truncation radius is shown in units of binary semi-major axis as a function of binary mass ratio for a zero-eccentricity binary (solid lines) and for $\ebin=0.5$ (dashed-lines). Both the truncation radius of the circumprimary and circumsecondary stars are displayed. We note that the binary eccentricity has a large impact on the truncation radius. For instance, for an equal mass binary, the disc truncates at $1/3\times \abin$ for zero-eccentricity, but it truncates at $0.15\times \abin$ for $\ebin=0.5$. A similar plot, limited to the circular case, is presented in \citet{Rosotti18}, where the authors fit curves to the data points originally provided by \citet{Papaloizou1977}.
For convenience, we provide an open source python script\footnote{\url{https://github.com/Aryy98/Circumstellar_disc_truncation_radius}} to compute the disc truncation following the outlined  prescription.

We emphasise that this method allows us to model the disc truncation either for the circumprimary or the circumsecondary disc, depending around which star we want to study planet formation. The code does not allow to simulate planet formation around the two stars simultaneously. Nevertheless, it is possible to study planet formation around each of the two stars separately, with different sets of simulations. This is particularly compelling when trying to reproduce observed systems where planets have been detected around the two components. There are currently seven known systems with planets around the primary and the secondary: Kepler-132 (A and B), WASP-94 (A and B), XO-2 (N and S), HD20782, HD20781, HD113131 and 55 Cancri \citep[][and references therein]{55Cnc_2025}.

\begin{table} [h]
    \caption{Table C.1 from \citet{Manara19}, which shows the best fit parameters for Eq. \eqref{eq:Rtrunc} for different values of $\mu$ and $\mathcal{R}$, both for circumprimary and circumsecondary disks.} 
    \label{tab:Manara2019}
    \centering
    \begin{tabular}{cccccc}
        \hline
        \hline
        \multicolumn{3}{c}{Circumprimary} & \multicolumn{3}{c}{Circumsecondary}\\
        \hline
        $\mathcal{R}$ & $b$ & $c$ & $\mathcal{R}$ & $b$ & $c$\\
        \hline
        \multicolumn{6}{c}{$\mu=0.1$}\\
        \hline
          $10^4$ & $-0.66$ & $0.84$ & $10^4$ & $-0.81$ & $0.98$\\
          $10^5$ & $-0.75$ & $0.68$ & $10^5$ & $-0.81$ & $0.80$\\
          $10^6$ & $-0.78$ & $0.56$ & $10^6$ & $-0.83$ & $0.69$\\
        \hline
        \multicolumn{6}{c}{$\mu=0.2$}\\
        \hline
          $10^4$ & $-0.72$ & $0.88$ & $10^4$ & $-0.81$ & $0.99$\\
          $10^5$ & $-0.78$ & $0.72$ & $10^5$ & $-0.82$ & $0.82$\\
          $10^6$ & $-0.80$ & $0.60$ & $10^6$ & $-0.83$ & $0.69$\\
        \hline
        \multicolumn{6}{c}{$\mu=0.3$}\\
        \hline
          $10^4$ & $-0.76$ & $0.92$ & $10^4$ & $-0.79$ & $0.97$\\
          $10^5$ & $-0.80$ & $0.75$ & $10^5$ & $-0.82$ & $0.81$\\
          $10^6$ & $-0.81$ & $0.63$ & $10^6$ & $-0.83$ & $0.69$\\
        \hline
        \multicolumn{6}{c}{$\mu=0.4$}\\
        \hline
          $10^4$ & $-0.77$ & $0.95$ & $10^4$ & $-0.80$ & $0.98$\\
          $10^5$ & $-0.81$ & $0.78$ & $10^5$ & $-0.82$ & $0.80$\\
          $10^6$ & $-0.82$ & $0.66$ & $10^6$ & $-0.83$ & $0.68$\\
        \hline
        \multicolumn{6}{c}{$\mu=0.5$}\\
        \hline
          $10^4$ & $-0.78$ & $0.94$ & $10^4$ & $-0.79$ & $0.95$\\
          $10^5$ & $-0.81$ & $0.78$ & $10^5$ & $-0.81$ & $0.78$\\
          $10^6$ & $-0.82$ & $0.66$ & $10^6$ & $-0.82$ & $0.66$\\
         \hline
    \end{tabular}
\end{table}

\subsection{Evolution of the gas disc}
We adopt the standard 1D axis-symmetric disc evolution model from the Bern Model, adapting it to the presence of the companion star by imposing that the disc outer edge is the tidal truncation radius given by Eqs.\ref{eq:Rtrunc}-\ref{eq:Rtrunc3}. 
The disc evolves by viscous accretion as well as by internal and external photoevaporation. The averaged gas surface density and viscosity at the disc midplane are used to solve the radial diffusion equation \citep{Pringle1981} with sink terms due to photoevaporation from the host star and planetary gas accretion. As in \citet{Zagaria2021}, we adopt a zero-flux boundary condition at the truncation radius:
\begin{equation}
\left\{
\begin{array}{l}
\displaystyle
\frac{\partial \Sigma_{\rm gas}}{\partial t}= \frac{3}{R}\frac{\partial}{\partial r} \left[ r^{1/2} \frac{\partial}{\partial r} \left( \nu \Sigma_{\rm gas} r^{1/2}  \right) \right] - \dot{\Sigma}_{\rm photo} (r) - \dot{\Sigma}_{\rm planet}(r), \\
 \quad  \quad  \quad  \quad  \quad  \quad   \quad  \quad  \quad  \quad  \quad  \quad  \quad  \quad  \quad  \quad  \quad  \quad  \, \text{if } r \leq R_{\rm trunc}, \\
\displaystyle \frac{\partial}{\partial r} (\nu \Sigma_{\text{g}} r^{1/2}) = 0 \quad \text{if } r = R_{\rm trunc}
\end{array}
\right.
\label{eq_gas-evol}
\end{equation}
where $t$ and $r$ are the temporal and radial coordinates, $\Sigma_{\rm gas}$ is the gas surface density, and $\nu= \alpha c_s \text{H}_{\text{g}} $ is the kinematic viscosity, given by the dimensionless parameter $\alpha$ \citep{SS73}, the local sound speed ($c_s$) and the disc's scale height ($H_{\rm g}$). 
$\dot{\Sigma}_{\rm photo} (r)$ is the sink term due to the internal and external photoevaporation. The internal photoevaporation follows \citet{Clarke2001}, while the  external photoevaporation is modelled using the far-ultraviolet prescription of \citet{Matsuyama2003}. We emphasise that in this work we have not modified the photoevaporation rates compared to the nominal NGPPS series \citep{Emsenhuber21}. We discuss in Appendix B how external photoevaporation affects the disc lifetimes and its limiting role in the case of a disc truncated by a close secondary companion.

\subsubsection*{Vertical structure}
The vertical structure of the disc is computed following the Bern Model in its version of \citet{Emsenhuber21}, which employs the semi-analytical approach from \citet{Nakamoto94} and \citet{HuesoGuillot05}. We adapt the equation for the disc's midplane temperature by adding a term corresponding to the tidal heating stemming from the torque exerted by the stellar companion ($Q_{\rm tidal}$ term), following \citet{Alexander11}. The disc midplane temperature is thus:
\begin{equation}
\sigma_{\rm} T^4_{\rm mid} = \frac{1}{2} \bigg(\frac{3}{8} \tau_R + \frac{1}{2\tau_P}\bigg)\dot{E} + \sigma_{\rm} T^4_{\rm S} + \bigg( 1+\frac{1}{2\tau_P}\bigg)\Sigma Q_{\rm tidal}
\label{mid_plane_temp}
\end{equation}
with $T_{\rm mid}$ the disc mid-plane temperature, $T_S$ the temperature due to the irradiation (see below), $\sigma$ the Stefan-Boltzmann constant,
$\tau_R$ and $\tau_P$ are the Rosseland and Planck mean optical depths
respectively, and $\dot{E}$ is the viscous dissipation rate. This formula
yields the mid-plane temperature both in the optically-thick (the
term with $\tau_R$) and optically-thin (the term with $\tau_P$) regimes. The computation of $\dot{E}$,  $\tau_R$  and $\tau_P$ is specified in Sect.3.2.1 of \citet{Emsenhuber21}.

The tidal heating term due to the torque exerted by the companion star is \citep{Alexander11}:
\begin{equation}
Q_{\rm tidal} = |\Omega_{\rm bin} - \Omega(r)| \Lambda(r) \Sigma(r) dr
\end{equation}
and 
\begin{equation}\label{Eq_lambda}
    \Lambda =- \frac{q^2 G M_1}{2r} \bigg(\frac{r}{\Delta_p}\bigg)^4
\end{equation}
where $\Delta_p = \rm{max}(H,|r-\abin|)$

The irradiation temperature $T_{\rm S}$ is calculated as:
\begin{align} \label{Eq_Ts}
    T^4_{\rm S} &= T^4_{\star,1} \bigg[\frac{2}{3\pi} \bigg(\frac{R_{\star,1}}{r}\bigg)^3 +  \frac{1}{2}\bigg(\frac{R_{\star,1}}{r}\bigg)^2 \frac{H}{r} \bigg(\frac{\partial {\rm ln} H}{\partial {\rm ln r}} -1 \bigg) \bigg] + T^4_{\rm irr, 1}  \\
   & + T^4_{\rm irr, 2} + T^4_{\rm min,disk}  \nonumber  
\end{align}
This expression is the analogous to Eq.5 from \citet{Emsenhuber21}, with quantities with sub-index "1" referring to the primary star, and with "2" to the secondary star. The difference with  Eq.5 from \citet{Emsenhuber21} is that we consider not only the direct irradiation term through the disc midplane from the primary $T^4_{\rm irr,1}$, but from the secondary as well ($T^4_{\rm irr,2}$).
The term $T^4_{\rm irr, 1}$ is given by \citep[][Eq.6]{Emsenhuber21}:
\begin{equation}
    T^4_{\rm irr, 1} = \frac{L_{\star,1}}{16 \pi r^2 \sigma} e^{-\tau_{\rm mid}}
\end{equation}
where $L_{\star,1}$ is the primary's luminosity, $\sigma$ the Stefan-Boltzmann constant, and $\tau_{\rm mid}$ the midplane's optical depth, given by $\tau_{\rm mid} = \int \kappa(r) \rho(r) dr$, with the integral performed from the disc inner edge until the location $r$ in the disc.

In the nominal Bern Model, $T_{\rm min,disk}$ from Eq.\ref{Eq_Ts} is the background temperature fixed at 10 K \citep{Emsenhuber21}. This term accounts for the heating by the surrounding environment (molecular cloud). In the presence of a stellar companion, beyond this "background temperature", there is also the irradiation from the secondary star, which we include directly as $T^4_{\rm irr, 2}$ in Eq.\ref{Eq_Ts}. We note that this term could be complex if one wants to compute it taking into account the disc and binary's geometry, as well as if there is a circumsecondary disc blocking the light stemming from the secondary into its way to the circumprimary disc. Thus, we adopt a very simple approach. We only consider a direct irradiation term from the secondary into the circumprimary disc, at the midplane, and considering only the equilibrium temperature that would emerge due to the irradiation from the secondary. Thus,
\begin{equation}
    T^4_{\rm irr, 2} = \frac{L_{\star,2}}{16 \pi r_2^2 \sigma} e^{-\tau'_{\rm mid}}
\end{equation}
where $r_2$ is the distance between the given location in the circumprimary disc and the secondary star, and $L_{\star,2}$ is the luminosity of the secondary star. For simplicity, we use the stellar mass-luminosity relation $L_{\star,2} \sim M^{3.5}_2$ (normalised for the present solar luminosity). The optical depth $\tau'_{\rm mid}$ is analogous to $\tau_{\rm mid}$ describe above, except that the integral is performed from the disc truncation radius until the disc inner edge. 
To compute $r_2$ we use an average distance as for the computation of $\Delta p$ in Eq.\ref{Eq_lambda}.
We note that this term starts to be non-negligible when the luminosity of the secondary is of the same order of the luminosity of the primary star (i.e, binary mass ratios close to 1).

\subsection{Evolution of the dust disc and pebble accretion}\label{sec_meth_vfrag}
The solid accretion is assumed to be dominated by pebbles. To compute realistic pebble accretion rates, it is essential to properly compute the pebbles' sizes \citep{Venturini20SE, Venturini20Letter}. For this, we compute the dust evolution by coagulation, fragmentation, drift, and ice sublimation at the ice line. We adopt the two-population model from \citep{Birnstiel12}, which describes the dust population by two dominant sizes: dust grains at a fixed size of $10^{-5}$ cm and pebbles of evolving sizes. Both dust and pebbles evolve embedded in the truncated gaseous discs presented in Sect.\ref{sec_disc_truncation}.
The two-population model was introduced in the Bern Model by \citet{Voelkel20}. Unlike \citet{Voelkel20}, where
a single pebble fragmentation velocity was used across the entire
disc, we adopt a fragmentation velocity in accordance with the pebbles' composition: v$_{\rm frag}$ = 1 m/s inside the ice line and v$_{\rm frag}$ = 10 m/s outside it \citep{Blum2018, Drazkowska17, Guilera20}.

For pebble accretion, we follow the prescription of \citet{JohansenLambrechts2017}, incorporating the effect of halting pebble accretion once a planet reaches the pebble isolation mass, as described in \cite{Lambrechts14}, and the effect of the planet eccentricity on the pebble accretion rate \citep[][eq.36]{JohansenLambrechts2017}.





\section{Results} \label{sec_results}
\subsection{Gas Disc evolution}\label{sec_gas-disc}
\begin{figure*}[h!]
    \centering
    \includegraphics[width=\linewidth]{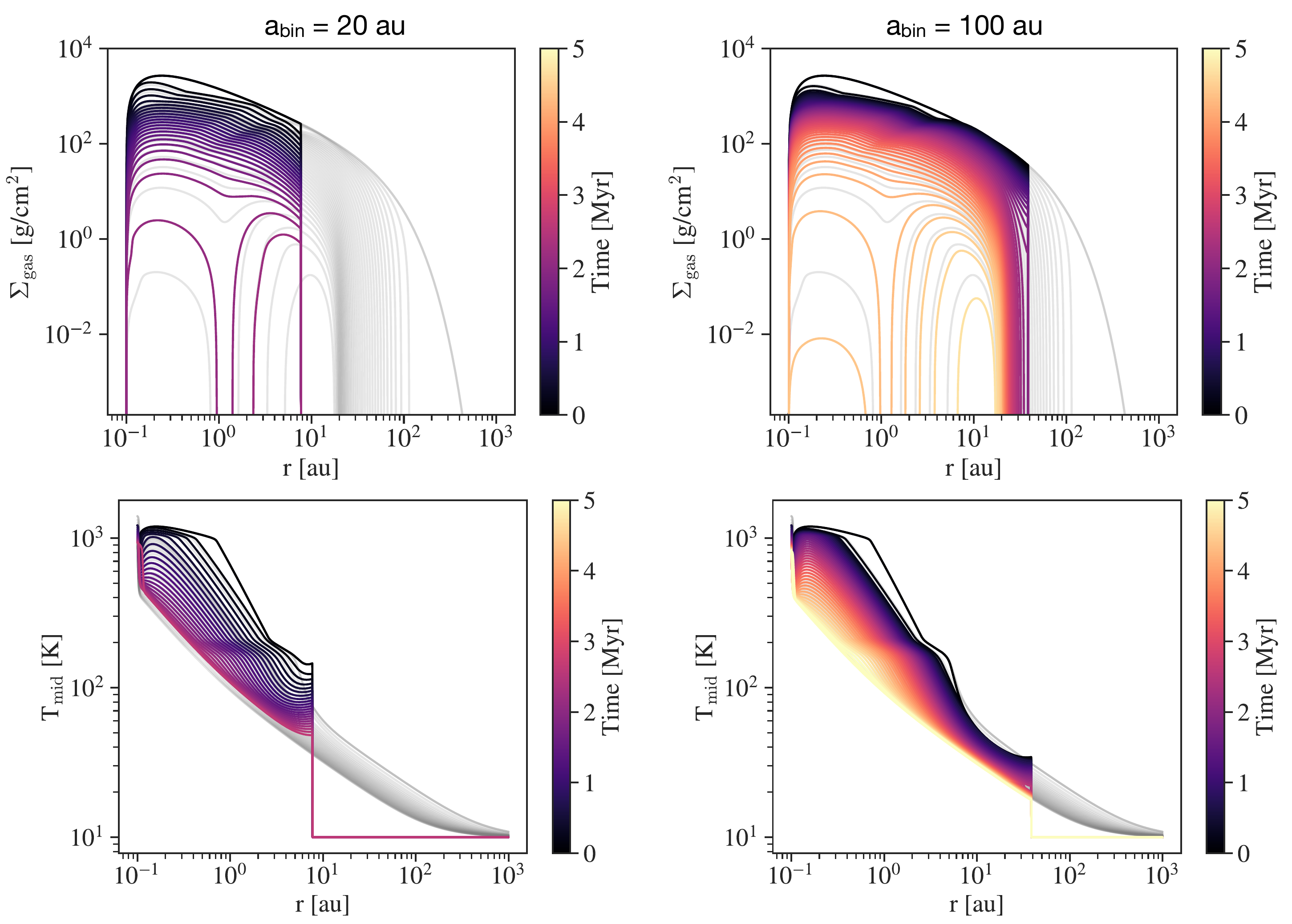}
    \vspace{-0.6cm}
    \caption{Evolution of the gas disc as a function of orbital distance from the primary, for the nominal disc with binary separations of $\abin=20$ au (left) and $\abin=100$ au (right). The grey background curves correspond to the single-star case. The disc's profiles are shown every $10^5$ years. Top panels display the evolution of the gas surface density, bottom panels the evolution of the disc midplane temperature.}
    \label{fig_gas_disc}
\end{figure*}

We first analyse the evolution of the gas disc for the nominal setup, defined by a disc with $\alpha=10^{-3}$, initial total disc mass before truncation of $M_{\rm d,0}= 0.1 \, \Msun$, and initial dust-to-gas ratio of 0.01. The binary parameters for this nominal setup are $M_1 = 1 \, \Msun$, $M_2 = 0.5 \, \Msun$, $\ebin = 0$ and variable $\abin$. All the simulations presented in this work assume coplanarity between the binary and the protoplanetary disk. The initial conditions are also displayed for convenience in the middle column of Table \ref{tab:initialconditions}, except that for the results of Sect.\ref{sec_gas-disc} and Sect.\ref{sec_dust_disc} we do not include the growth of any planet, since we are interested in understanding first the disc evolution alone. Figure \ref{fig_gas_disc} shows the evolution of the surface density of gas (top panels) and of the midplane temperature as a function of radial distance to the primary, for the cases where $\abin=20$ au (left panels) and $\abin=100$ au (right panels). The single-star case is shown for reference in the background. The disc disappears at 3.5 Myr for the single star case, at 2.1 Myr for $\abin=20$ au, and at 4.3 Myr for $\abin=100$ au. 
The gap in the gas surface density at a few au that appears close to the disc dissipation is due to the effect of the internal photoevaporation, which more easily removes gas at intermediate distances \citep[see, e.g.][]{Venturini20SE}.

\begin{table} [h]
    \caption{Initial conditions for the simulations presented in Sec. \ref{sec_planet-growth} (left column) and in Sec. \ref{sec:parameter_study} (right column).}
    \label{tab:initialconditions}
    \centering
    \begin{tabular}{l|l|l}
        \hline
        \hline
        & simulations \ref{sec_planet-growth} & simulations \ref{sec:parameter_study} \\
        \hline
        \multicolumn{3}{c}{Disc parameters}\\
        \hline
          $M_{\text{d,0}}$ [$\Msun$]& 0.1 & Log-$\mathcal{U}$[0.001,0.1] \\
          $f_{\rm D/G}$ & 0.01 & Log-$\mathcal{U}$[0.0056,0.03162] \\
          $\alpha$ & 0.001 & Log-$\mathcal{U}$[$10^{-4},10^{-3}$] \\
          $r_c$ [au] & 50 & Log-$\mathcal{U}$[10,200]  \\
          \hline
        \multicolumn{3}{c}{Binary parameters}\\
        \hline
          $M_1$ [$\Msun$] & 1 & 1 \\
          $M_2$ [$\Msun$] & 0.5  & $\mathcal{U}$[0.1, 1]\\
          $\ebin$\   & 0 & $\mathcal{U}$[0, 0.9]\\
          $\abin$\  [au] & 20, 50, 75, 100, 300 & Log-$\mathcal{U}$[10, 1000]\\
        \hline
        \multicolumn{3}{c}{Planet parameters}\\
        \hline
          $N_{\text{p}}$ & 1 & 1 \\
          $a_{\text{p,0}}$ & 5, 20 [au] & Log-$\mathcal{U}$[1 au, $R_{\text{trunc}}$ - 1 au]\\
          $\kappa$ & 0.01$\times$BL94 & 0.01$\times$BL94 \\
         \hline
    \end{tabular}
    \tablefoot{BL94: \citet{BL94}, $M_{\text{d,0}}$ is the total disc mass before truncation.}
\end{table}

Regarding the evolution of the disc's midplane temperature, we note that the $\abin=100$ au case is very similar to the single star case. On the other hand, for $\abin=20$ au there is a considerable temperature increase close to the truncation radius when comparing binary- with single-star case (increase in temperature from 70 K to 140 K for the first time-step). This is the effect of the tidal heating from the stellar companion, and it is only noticeable for very close binaries and mainly at the beginning of the evolution. 

The change in the temperature profile at r$\sim$0.4-5 au is due to the water iceline for all the cases. Indeed, the presence of icy grains increases the disc opacity (modelled with the BL94 opacities), abating the drop in temperature with orbital distance. This change in slope moves inwards with time with the inner movement of the iceline, as the disc cools down.

\begin{table}[]
    \centering
    \begin{tabular}{c|c|c|c|c}
    \hline
    \hline
      $\abin$ & $\Rtrunc$  & M$_{\rm g,0}$ & M$_{\rm s,0}$ & $\tau_{\rm disc}$ \\
        \rm{[au]}& [au] &  [$\Msun$]& [\ME]& [Myr]  \\
       \hline
       20  & 7.7 & $9.5\times10^{-3}$& 31.57 & 2.1 \\
       50 & 19.3 & $2.5\times10^{-2}$& 84.54 & 6.3 \\
       75 &  26.9 & $3.7\times10^{-2}$& 123.41 & 4.8 \\
       100 & 38.5 & $4.7\times10^{-2}$& 157.13 & 4.3 \\
       300 & 116 & $8.5\times10^{-2}$& 282.14 & 3.9 \\
       single-star & - & $9.3\times10^{-2}$& 307.68 & 3.9 \\
       \hline
    \end{tabular}
    \vspace{0.2cm}
    \caption{Disc properties for different binary separations for $\ebin=0$. From left to right: binary semi-major axis, truncation radius, initial gas disc mass after truncation, initial solid disc mass after truncation, and disc lifetime. The initial solid disc mass after truncation is computed as the total disc mass after truncation times the initial dust-to-gas ratio ($f_{\rm D/G}$ in Table \ref{tab:initialconditions}).} 
    \label{tab_gas-disc_No-planets}
\end{table}

Table \ref{tab_gas-disc_No-planets} displays the discs' characteristics for the cases presented in Fig.\ref{fig_gas_disc}, as well as for other binary separations.
We note that the gas disc lifetimes do not always decrease when reducing the binary separation, as one would intuitively expect given that the more truncated the disc, the less massive it is, and thus, the easier it should be to remove the gas. The discs that we are modelling, which stem from the nominal set up of the Bern Model, are discs whose evolution is driven by external photoevaporation. External photoevaporation produces discs that evolve "outside-in" \citep[e.g.][]{Coleman2022}, i.e, external photoevaporation efficiently removes material from the outer edge, dragging along gas from the inner orbits until R$\approx$20 au. When a disc gets truncated within this radius, external photoevaporation cannot affect the disc evolution, rendering longer disc lifetimes. We analyse this effect in more depth in Appendix \ref{App_externalFoto}.

In any case, we find that the limiting factor for planetary growth is not the gas disc lifetime but the \textit{pebbles' disc lifetime}. This was also reported in \citet{Zagaria2021}. Indeed, the pebbles disappear so quickly in truncated discs due to the fast radial drift and lack of a reservoir of dust further out, that the planetary seeds are quickly drained from solid material, preventing core growth. We analyse this in more detail below. 

\begin{figure*}[h!]
    \centering
    \includegraphics[width=\linewidth]{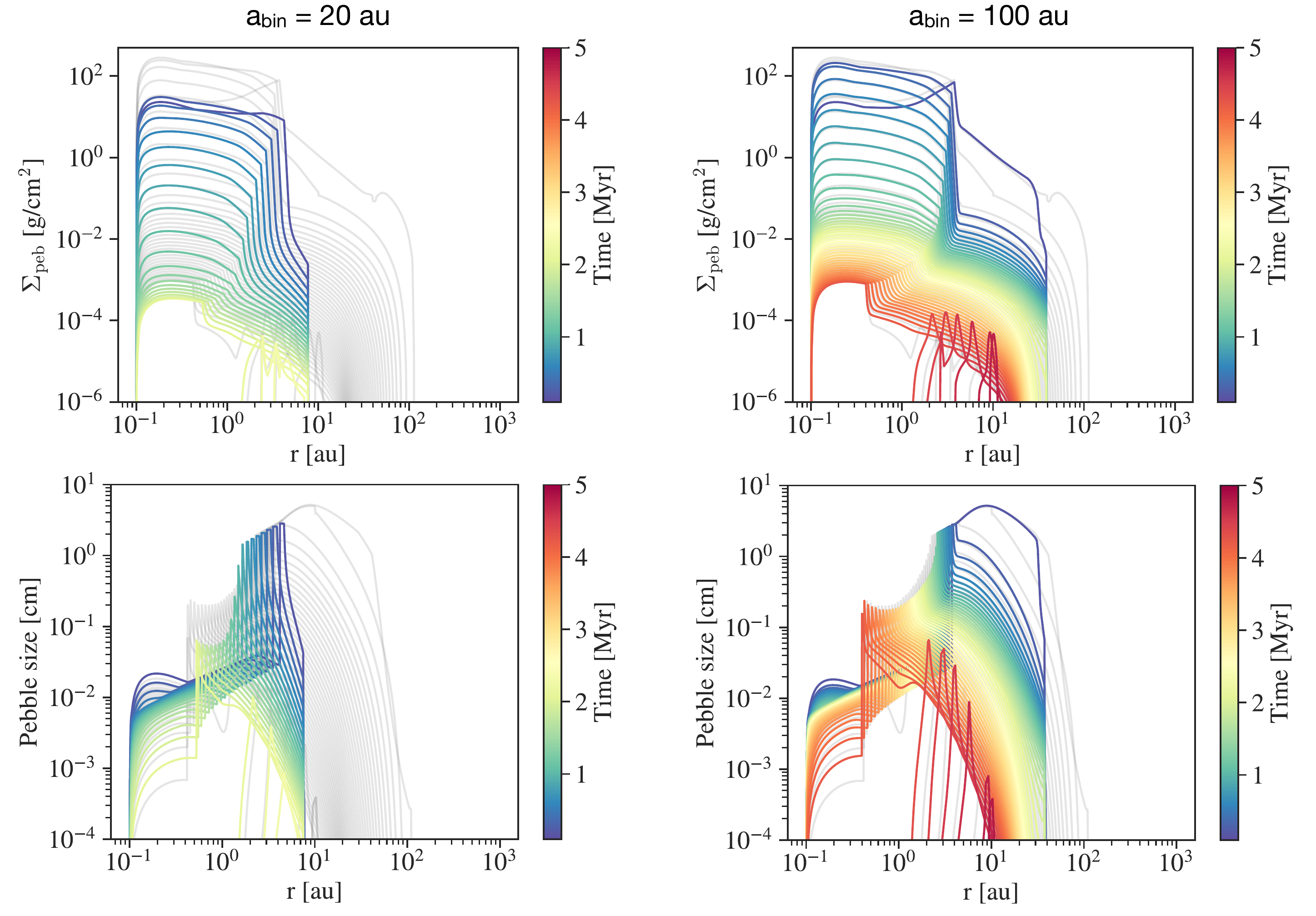} 
    \vspace{-0.6cm}
    \caption{Evolution of the pebbles' disc as function of orbital distance from the primary, for the nominal disc with binary separations of $\abin=20$ au (left) and $\abin=100$ au (right). The grey background curves correspond to the single-star case. The disc's profiles are shown every $10^5$ years. The top panels indicate the evolution of the pebbles' surface density, while the bottom panels show the evolution of the pebbles' sizes. The abrupt change of the pebbles' size at r$\sim$0.4-5 au for all the profiles and times corresponds to the location of the water iceline. This affects as well the profiles of pebbles' surface density.}
    \label{fig_pebble_disc}
\end{figure*}

\subsection{Dust Disc evolution}\label{sec_dust_disc}
Figure \ref{fig_pebble_disc} shows the evolution of the surface density of pebbles (top panels) and of the pebbles' size along the disc, for the nominal disc with $\abin=20$ au (left) and $\abin=100$ au (right). The grey background lines correspond to the single-star case. For all the cases, the abrupt increase on the pebbles' size when moving radially outwards, visible at r$\sim0.4-5$ au, corresponds to the presence of the iceline. Icy pebbles have larger fragmentation velocities (see Sect.\ref{sec_meth_vfrag}), and thus can grow to larger sizes compared to rocky pebbles. The change of the pebbles' size affect as well the radial drift, with smaller rocky pebbles moving inwards slower. This creates the well-known pile-up of pebbles inside the iceline \citep[e.g.][]{Drazkowska16}, visible in the profiles of pebbles' surface density of Fig. \ref{fig_pebble_disc}.

On the other hand, the peaks of pebbles towards the end of the disc dissipation emerge as a response to the gap in the gas midplane, carved by internal photoevaporation as mentioned in Sect.\ref{sec_gas-disc}.
Nevertheless, it is particularly interesting to analyse the differences in the pebbles' evolution within the first timesteps, because the core growth by pebble accretion happens very quickly, in timescales of $\sim10^5$ years (see next section, Fig.\ref{fig_growth}). When comparing the first time-snapshot of Fig. \ref{fig_pebble_disc} (at time=$10^5$ years), we note that at $r=5$ au the surface density of pebbles is $\Sigma_{\rm peb} = 10^{-2}$ g/cm$^2$, while for the single star case it is  $\Sigma_{\rm peb} = 4$ g/cm$^2$, that is, a reduction by a factor 400 for  $\abin=20$ au compared to the single-star case after the first $10^5$ years of disc evolution. This sharp drop in pebbles surface density is a consequence of the disc truncation and the loss of the outer reservoir of dust.
As intuition dictates, the effect is less dramatic the larger the binary separation. Indeed, for $\abin=100$ au, the effect is noticeable at the second time-snapshot (time=$2\times10^5$ years), where the surface density of pebbles for the truncated disc is $\Sigma_{\rm peb} = 0.04$ g/cm$^2$ compared to $\Sigma_{\rm peb} = 0.94$ g/cm$^2$ of the single-star case at $r=5$ au (a drop by a factor 23.5 in this case).

The halting of the pebbles' supply also affects the pebbles' growth: the more truncated the disc, the smaller the pebbles at a given time, as illustrated in the bottom panels of Fig.\ref{fig_pebble_disc}. For the case of $\abin=20$ au, the situation is dramatic outside the ice line as we move towards $\Rtrunc$, where the pebble size drops by a factor of 6 at $r=5$ au and by a factor of 35 at $r=7$ au compared to the single-star case withinin the first snapshot of the displayed evolution (time=$10^5$ years). Both the surface density of pebbles and the pebbles' sizes affect the rate of pebble accretion to build the cores \citep{JohansenLambrechts2017}. Thus, both effects play a detrimental role in the growth of the planetary cores, as we analyse in the next section.


\subsection{Planetary growth}\label{sec_planet-growth}
\begin{figure}[h!]
    \centering
    \includegraphics[width=\linewidth]{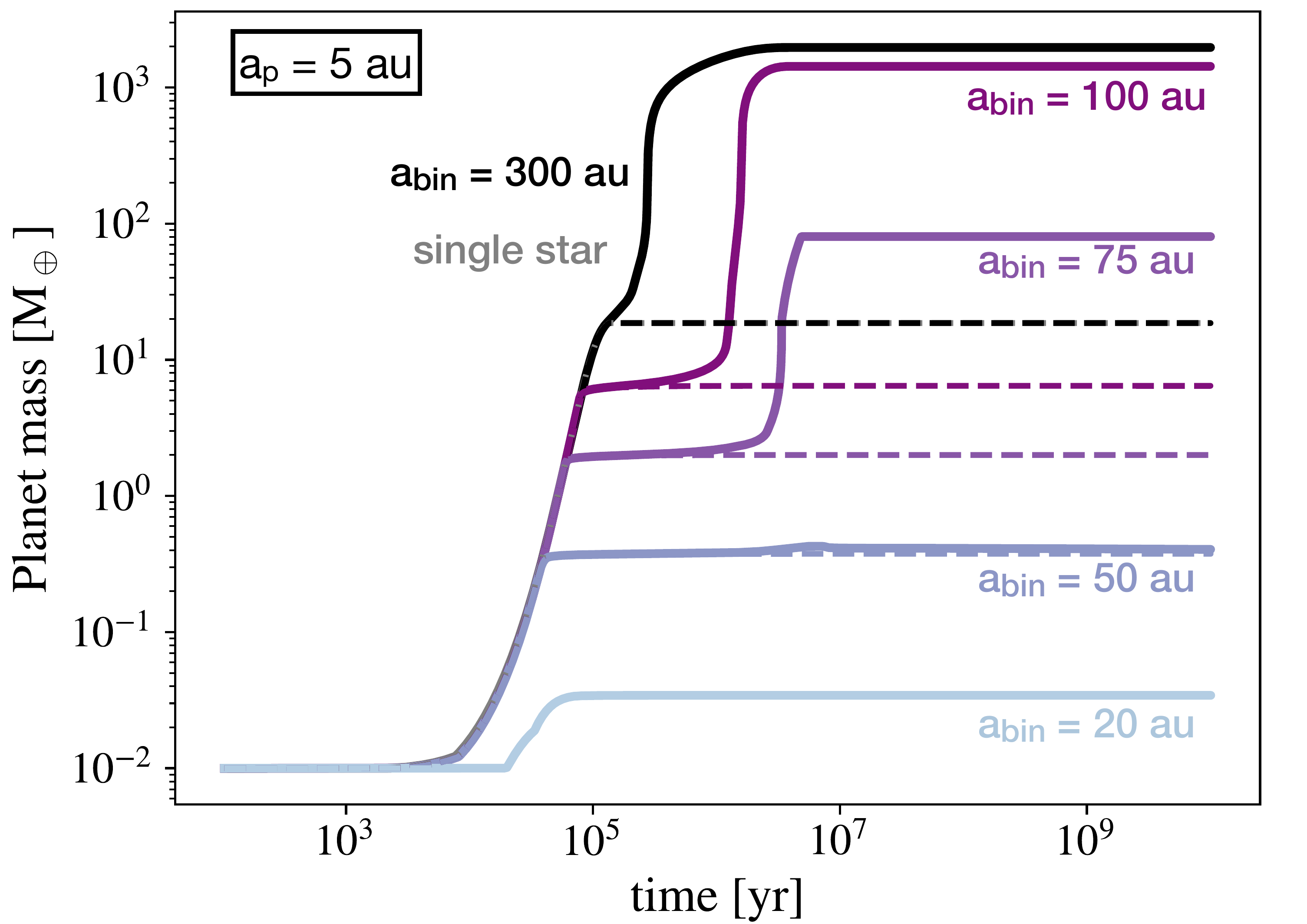} 
    \includegraphics[width=\linewidth]{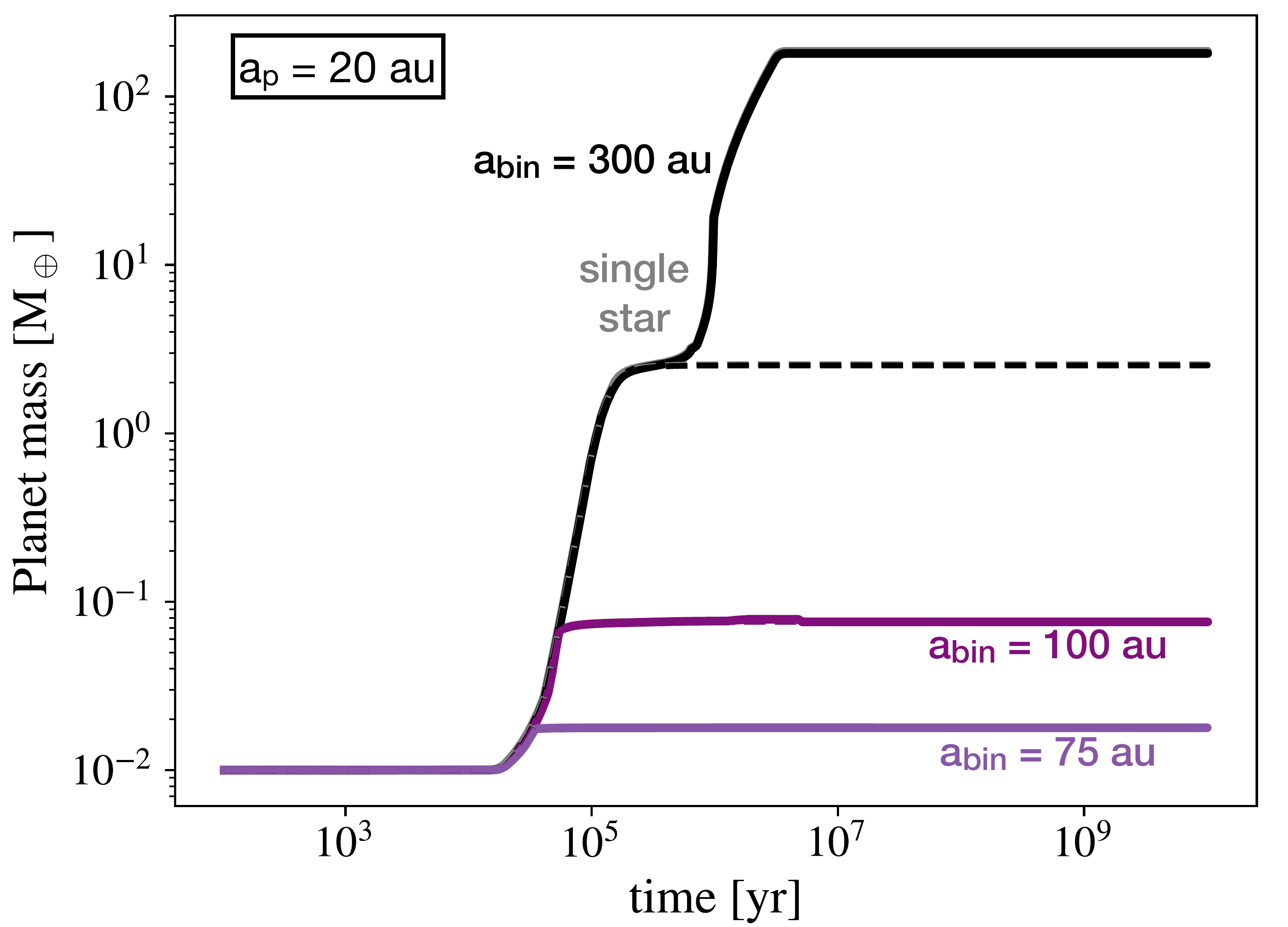}
    \vspace{-0.5cm}
    \caption{In-situ planet growth by pebble accretion at $a_p=5$~au (top panel) and $a_p =20$~au (bottom panel). The solid lines indicate the evolution of the total planet mass, while the dashed lines the evolution of the core mass. The different colours correspond to different binary separations, as indicated in the labels. After the final planet mass is reached during the disc lifetime (a few million years), the planet evolves during giga-years by cooling and contraction.}
    \label{fig_growth}
\end{figure}

In this section we analyse how a single moon-mass embryo grows by pebble and gas accretion at fixed locations (no migration), in discs truncated by the tidal effect of a secondary star. We also show the single-star cases for reference. Figure \ref{fig_growth} shows the growth at $a_p = 5$ au and $a_p = 20$ au, for different binary separations and for the single-star case. We note that for all cases, the core completes its formation in a timescale of $10^5$ years, while gas keeps being accreted until the disc dissipates (a few million years, see the corresponding discs' characteristics in Table \ref{tab_gas-disc_No-planets}). After disc dissipation, the planet evolves by cooling and contracting at constant mass during giga-years.

For $a_p = 5$ au (Figure \ref{fig_growth}, top panel) we note that a binary separation of 300 au produces the same planet as the single star case, and that we have to decrease the binary separation below 100 au to start to notice an effect on the planetary growth. 
It is worth to recall that for all the shown cases, the disc is truncated at $\Rtrunc \approx 0.4 \, \abin$ (see fig. \ref{fig_truncation}.) This is why for $a_p = 20$ au we cannot run cases with $\abin<50$~au, because the planet would be beyond the disc outer edge.

For the planet growing at 5 au, the most dramatic effect on the planet's growth is observed in the transition from $\abin=75$ to $\abin=50$ au, where the planet changes from reaching a mass of 80 \ME for $\abin=75$ to 0.4 \ME for $\abin=50$~au.
For a planet located at 20 au, the strong drop in planet growth occurs between $\abin=300$ and $\abin=100$ au, where it transitions from a $\sim200$ \ME planet for $\abin=300$ to an 0.5 Mars object for $\abin=100$ au. The detrimental effect on the planet growth for binary separations below approximately 50 au is thus evident in these examples. As we outlined in Sect.\ref{sec_results}, this is not related to the gas disc lifetime but to the \textit{dust} disc lifetime. We note that for all our cases, when the surface density of pebbles drops below $\Sigma_{\rm peb}\leq 0.01$~g/cm$^2$, the pebble sizes reduce to values lower than $a_{\rm peb}\leq$0.5 cm (beyond the iceline) and the pebble accretion rates become negligible ($\dot{M}_{\rm peb} \leq 10^{-6}$ \ME/yr), halting planetary growth. Thus, the pebbles that are useful for building the cores are long gone by the time the discs are 1 Myr (see again the evolution of $\Sigma_{\rm peb}$ and of the pebble sizes along the discs in Fig.\ref{fig_pebble_disc}). 

When trying to quantify the exact transition between forming a giant planet versus a sub-Earth-mass object for different binary parameters, we note that this depends on the embryo's initial location over the truncation radius, and that the most clear variable impacting this is the initial mass of solids in the disc. We show this in the following section.


\subsection{Parameter study}\label{sec:parameter_study}
To investigate how the reduced material supply caused by disc truncation affects the planet growth across a variety of binary configurations, we carry out a grid of 5000 simulations. We fix the mass of the primary star, $M_1$, at 1 $\Msun$ and randomly draw the secondary star's mass, $M_2$, between 0.1 and 1 $\Msun$, to sample a broad range of binary mass ratios $q$. The binary separation $\abin$ and eccentricity $\ebin$ are randomly sampled within the ranges 10- 1000 au, and 0-0.9, respectively. In each simulation, we model the in-situ growth of a single moon-mass embryo (i.e., $M_\text{p}=10^{-2}$ $\Mearth$), placed at a location $a_{\text{p,0}}$ randomly chosen between 1 au and the location of the truncation radius $R_{\rm trunc}$ minus 1 au. We also sample the initial disc mass, $M_{\text{d,0}}$, between $10^{-3}$ $\Msun$ and 0.1 $\Msun$, the dust-to-gas ratio, $f_{\rm D/G}$, between 0.0056 and 0.03162, the disc viscosity, $\alpha$, between $10^{-4}$ and $10^{-3}$ and the disc characteristic radius, $r_c$, between 10 au and 200 au. These chosen ranges are physically motivated by observational data of single-star discs and they were used by \citet{Emsenhuber21_II} and \citet{Weder2023}. All these parameters are sampled with a uniform or log-uniform distribution and we summarize these initial conditions in the right column of Table \ref{tab:initialconditions}.

\begin{figure}[h!]
    \centering
    \includegraphics[width=\linewidth]{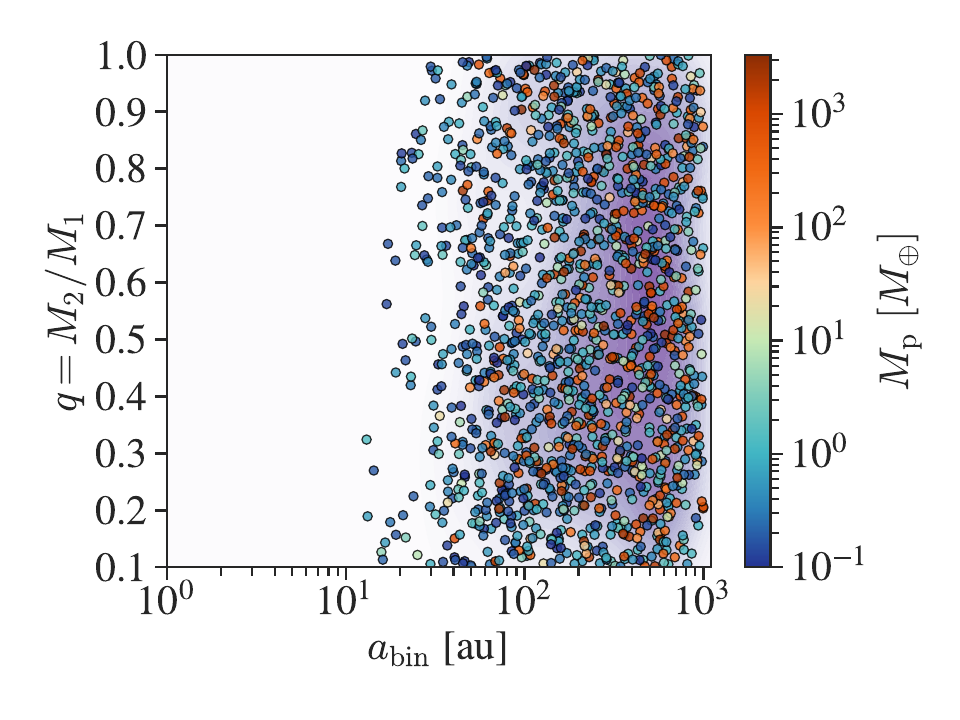} 
    \includegraphics[width=\linewidth]{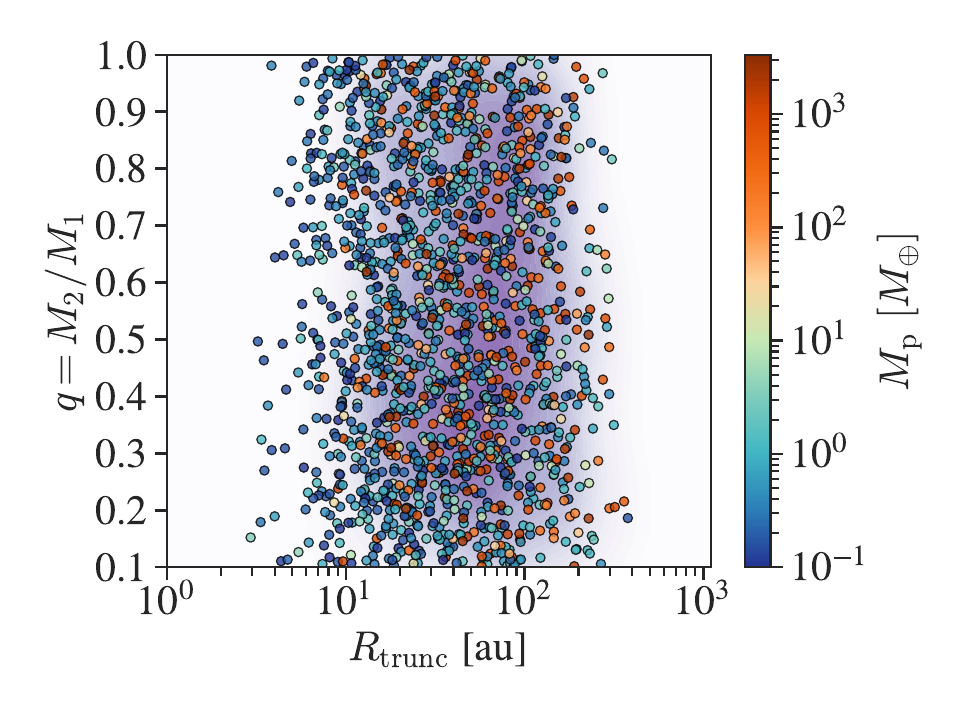}
    \vspace{-0.8cm}
    \caption{Binary mass ratio vs. binary separation (top panel) and binary mass ratio vs. disc truncation radius (bottom panel) for systems that formed a planet at least more massive than Mars. The colorbar indicates the final planet mass. Shaded contour regions represent areas of highest number planet density, derived from a two-dimensional kernel density estimation (KDE) performed on planets with masses greater than 10 $\Mearth$. The lack of planets for $\Rtrunc$>400 au stems from the choice of the upper limit on $\abin=1000$ au (see Table \ref{tab:initialconditions}).}
    \label{fig_param_study1}
\end{figure}

\begin{figure}
    \centering
    \includegraphics[width=1\linewidth]{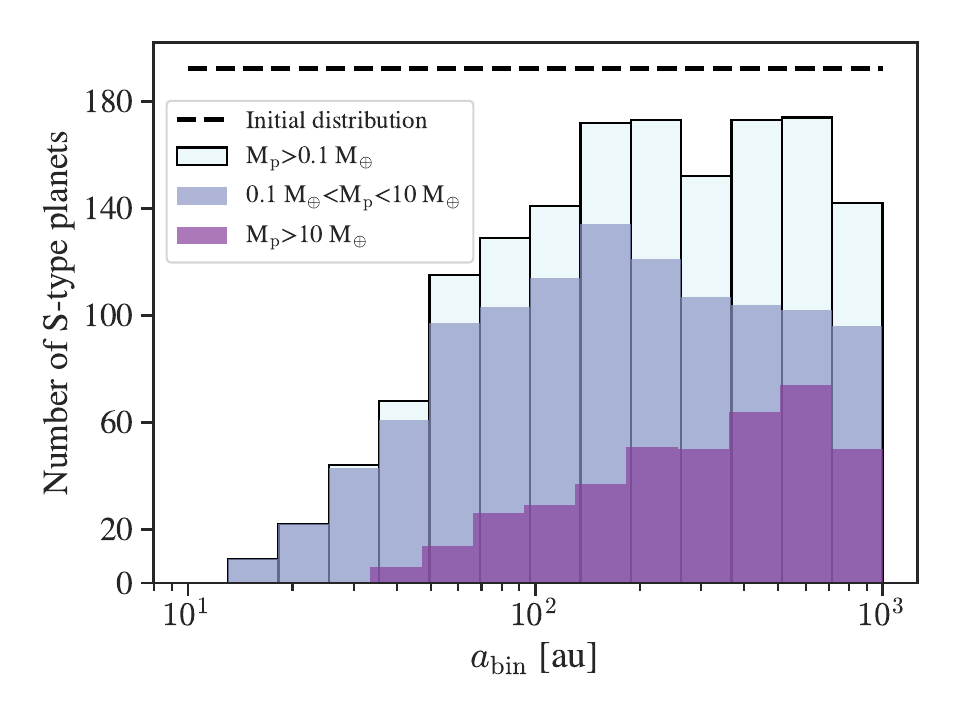}
    \vspace{-0.6cm}
    \caption{Number of formed S-type planets as a function of binary semi-major axis ($\abin$) from the grid of 5000 simulations presented in Sect.\ref{sec:parameter_study}. The black dashed-line illustrates the log-uniform distribution of $\abin$ adopted as the initial condition (multiplied by 0.5 for better visualisation). The different colours of the bars correspond to different ranges of planet mass, indicated in the legend of the figure. The simulations assume one embryo per disc and do not include the gravitational perturbation from the secondary, neither orbital migration.}
    \label{fig:grid_histo}
\end{figure}

\begin{figure}[h!]
    \centering
    \includegraphics[width=\linewidth]{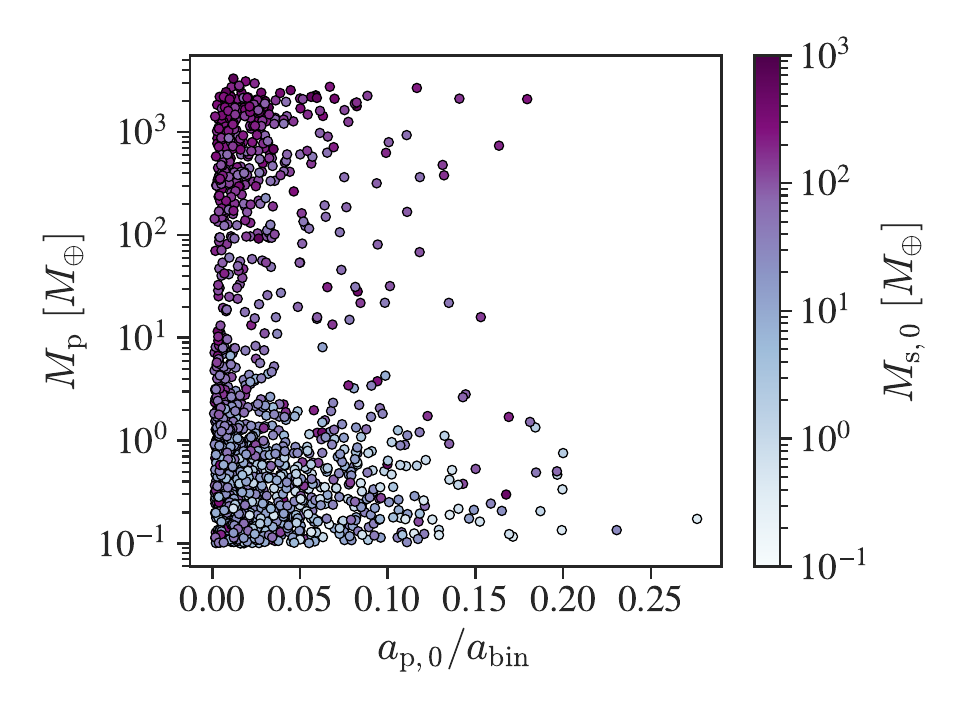} 
    \includegraphics[width=\linewidth]{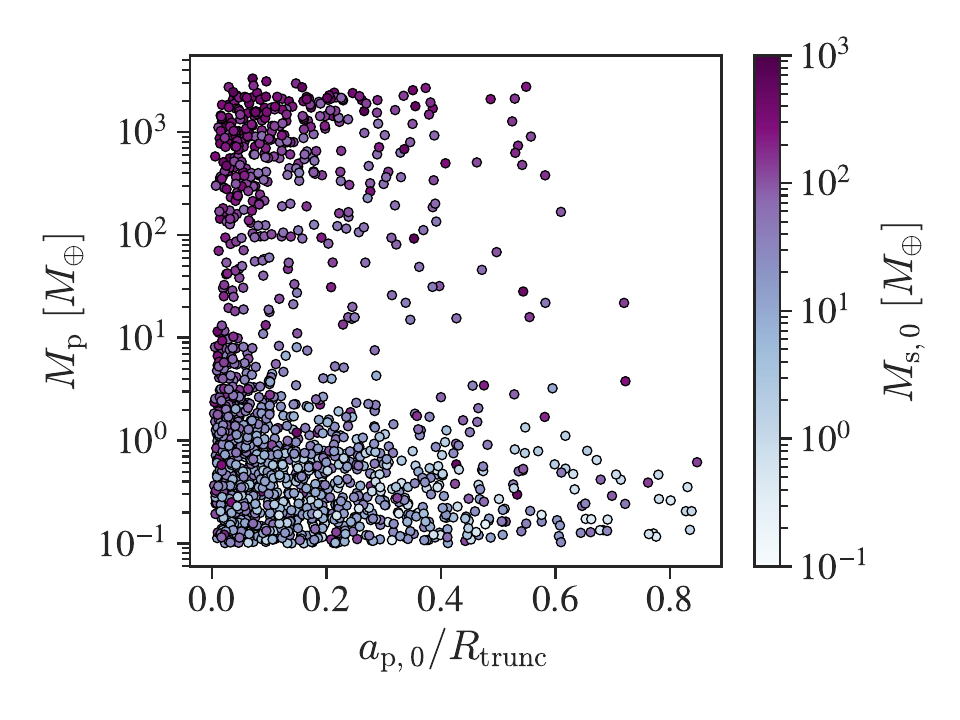}
    \vspace{-0.6cm}
    \caption{Initial planet location relative to the binary separation, $a_{\text{p,0}}/a_\text{bin}$, vs. final planet mass (top panel) and initial planet location relative to the disc truncation radius, $a_{\text{p,0}}/R_\text{trunc}$, vs. final planet mass (bottom panel) for systems that formed a planet at least more massive than Mars. The colour-bar indicates the initial mass of solids in the disc after truncation. Note the different x-axis scale in each plot.}
    \label{fig_param_study2}
\end{figure}

In Figure \ref{fig_param_study1}, we show the output of the planet formation simulations as a function of the binary mass ratio and binary separation (top panel) or truncation radius (bottom panel). In both panels, each circle is a planet, for which the final planet mass is indicated in the colour-bar. For better visualisation, we display only systems that formed planets more massive than Mars. The shaded purple contours mark areas of highest number density for planets with masses greater than 10 $\Mearth$. 

An important science question to address is: which type of planets do we form for different binary separations? We find that we form planets more massive than Mars practically everywhere, for all $\abin > 13$ au, when it is assumed that each log-bin of binary separation in the range $10<\abin<1000$ au is equally likely to produce planets. Nevertheless, we note that for planets more massive than 10 $\Mearth$, these form for $\abin>33.5$ au, while giant planets with $\Mp>100$ \ME form for $\abin>40.6$ au. These results hold irrespective of the binary mass ratio ($q$), and for all the binary eccentricities considered ($\ebin<0.9$). The are some specific trends in the final planet masses that depend on $\ebin$, but we analyse this effect in detail in Paper II. 

Another relevant aspect to analyse is the suppression of planet formation for decreasing binary separations. How does this operate when considering only the effect of tidal disc truncation? The reply is in Fig. \ref{fig:grid_histo}. It shows that when considering all the planets more massive than Mars, their formation becomes steadily suppressed for $\abin<160$~au. On the other hand, for planets more massive than 10 $\Mearth$, their suppression is noticeable for $\abin<600$ au. It is interesting to see that the number of small S-type planets drops mildly for $\abin>160$ au, while the number of giant planets keep rising for increasing $\abin$, up to $\abin~600$ au. The larger the binary separation, the more extended the discs, and thus, cores can accrete more solids, which leads to the formation of more giants (and thus less small planets). Thus, our results suggest that the answer to the question of planet formation suppression in binaries depends on the planet mass range that one wants to analyse: giant planet formation gets suppressed at larger binary separations compared to the formation of small planets, aspect that could explain the different findings of \citet{Lester2021} and \citet{Thebault_catlog-2025}.

As a word of caution, it is important to emphasise that these results hold under the assumption of an initial log-uniform distribution in $\abin$, which is not the observed distribution of binary separations in field binaries \citep{Raghavan10}. The purpose of this exercise is simply to analyse the effect of tidal disc truncation alone on the suppression of the formation of (single-embryo) S-type planets, when assuming all binary separations to be equally likely to produce planets. In addition, the gravitational perturbation from the secondary star is not included in these calculations, neither orbital migration. Thus, these numbers should not be directly compared to observations. We will present proper occurrence rates in a future population synthesis study (Paper III, Nigioni et al. in prep.)


As discussed in Section \ref{sec_planet-growth}, S-type planet formation depends not only on the binary properties but also on the planet’s location. In particular, planets more massive than Mars form only if located within 28\% of the binary separation, with 75\% of them being concentrated at only 4\% of the $\abin$ (Fig. \ref{fig_param_study2}, top panel). In terms of disc truncation, planets more massive than Mars locate at $a_{p,0}/\Rtrunc < 0.85$, with 75\% of them being concentrated at the innermost 22\% of the truncated disc (Fig. \ref{fig_param_study2}, bottom panel).
When focusing on more massive planets, we note that they are a bit closer-in: for $\Mp>10$ $\Mearth$, $a_{p,0}/\Rtrunc \lesssim 0.71$ and for $\Mp>100$, $a_{p,0}/\Rtrunc \lesssim 0.61$. Indeed, we note that in the illustrative examples of Fig.\ref{fig_growth}, for the formation location of $a_{p,0}=5$ au (top panel), the giant planet that forms for $\abin=75$ au has $a_{p,0}/\Rtrunc = 0.19$, while for the one forming at $a_{p,0}=20$ au, and $\abin=300$ au, $a_{p,0}/\Rtrunc = 0.172$ (disc truncation values taken from Table \ref{tab_gas-disc_No-planets}). 
When looking at the different variables that we analyse to understand the output of planet formation (see Figs.\ref{fig_param_study1} and \ref{fig_param_study2}), the initial mass of solids in the disc seems to be the most clear quantity influencing the final planet mass (Fig. \ref{fig_param_study2}). This a fundamental result of the core accretion model, also reported in the past for single stars \citep{Lokesh23}. For S-type planets the initial mass of solids in the disc is directly linked to the truncation radius, which depends linearly on $\abin$ (see Eq.~\ref{eq:Rtrunc}).

\section{Discussion} \label{sec_discussion}
\subsection{On the non-formation of $\gamma$--Cephei-like planets}
One of the best known cases of an S-type binary is $\gamma$-Cephei, an extreme S-type system with $M_1 \approx 1.3 \, \Msun$, $M_2 \approx\, 0.33 \Msun$ (binary mass ratio of q$\approx$0.26), a binary separation of $\abin\approx$ 20~au, binary eccentricity of $\ebin\sim$0.4, hosting a 6.6 Jupiter-mass planet orbiting the primary star at 2~au \citep{Hatzes2003, Knudstrup23}.
The disc of $\gamma$-Cephei should have been truncated at $\Rtrunc\approx4.4$ au (for $\alpha=10^{-3}$), according to the prescription presented in Sect.\ref{sec_methods}.
Different works have explored the formation, dynamical evolution, and long-term stability of $\gamma$-Cephei, generally agreeing that explaining the formation of its giant planet is very challenging \citep{Thebault2004,KleyNelson08,Jang-Condell2008,MullerKley2012,Jordan21}. The main difficulty for reproducing the $\gamma$-Cephei planet is the severe disc truncation by the eccentric stellar companion, which notably reduces the available mass of gas and dust to form the giant planet \citep{KleyNelson08,Zagaria2021}.
Hydrodynamical simulations suggest that the circumprimary disc can develop a significant eccentricity \citep{Kley2008,Marzari2009,Marzari2012,MullerKley2012}, which, combined with gravitational perturbations from the secondary star, could increase planetesimal relative velocities and lead to destructive rather than accretional collisions \citep{Paardekooper2008}. However, \citet{Beauge2010} found that for high disc eccentricities, disc precession can instead reduce the velocity dispersion between different-size planetesimals, favouring accretional collisions in the outer disc regions.

The formation of planets in binary systems, particularly in systems like $\gamma$-Cephei, has not yet been explored in the context of pebble accretion. In this first work, we address this gap by modelling planet formation under this paradigm. While this first approach simplifies certain aspects, such as the omission of detailed binary dynamics and planet migration, it provides valuable insights. Our results, especially those presented in Sect.\ref{sec:parameter_study}, show that pebble accretion, on its own, can form a gas giant ($\Mp>100 \, \Mearth$) for $\abin\gtrsim40$ au and $\Rtrunc\gtrsim7$ au. This means that with the current model we cannot strictly form a $\gamma$-Cephei-like planet. Nevertheless, it seems we are not far from forming a giant planet with the expected truncation radius of $\gamma$-Cephei. In fact, some of our simulations produce cores of 2-5 $\Mearth$ for $\Rtrunc \approx 2-7 $ au. With a bit different history of solid accretion or envelope opacity, such cores could enter into the runaway gas phase. Furthermore, the mass of the primary in $\gamma$-Cephei is 30\% larger than solar, meaning that more solids could have been present in the disc of $\gamma$-Cephei compared to the values considered in our simulations (solar-mass primaries). 
Thus, our simulations offer a starting point for a deeper exploration of the initial conditions and physical parameters that could lead to the formation of $\gamma$-Cephei Ab. We will address this in the future. 

Another potential solution to overcome the $\gamma$-Cephei problem has been recently proposed by \citet{MarzariDangelo2025}. The authors suggest that the system may have hosted an extended circumbinary disc, a remnant of the stellar formation process, that could have acted as a reservoir that supplied both gas and solids to the circumprimary disc. Through high-resolution hydrodynamical simulations, the authors showed that in a $\gamma$-Cephei like system, gas can be supplied by the outer circumbinary disc to the circumprimary disc, extending its lifetime for up to $\sim$3 Myr ($\sim$3 times the lifetime of the same disc in isolation). Moreover, \citet{MarzariDangelo2025} showed that solids can also be transported to the circumprimary disc. 
This mechanism could thereby have important implications for planet formation in close binaries, and we will examine it in a follow-up study (Ronco et al. in prep.). Extremely tight S-type binaries might undergo a different formation path as suggested by \citet{MarzariDangelo2025}, albeit more work is needed to quantify the binary parameters for which such scenario might work.

Finally, another formation pathway for $\gamma$-Cephei Ab could be gravitational instability (GI), which proposes that planets form through the rapid collapse of gas into self-gravitating clumps in sufficiently massive protoplanetary discs. For fragmentation to occur, however, the disc must cool efficiently so that self-gravity overcomes pressure support, a condition that is generally met only at large distances from the star \citep{Boss1998,Boss2002,Rafikov2005}. Thus, invoking GI as the formation pathway for $\gamma$-Cephei Ab (or for planets in relatively close binaries in general) could be challenging for the physical properties of tidally truncated discs. As we showed, discs in close binary systems are expected to be significantly less massive than around single stars due to the truncation of their outer regions, preventing them from reaching higher surface densities necessary for fragmentation to occur. In addition, these truncated discs are typically hotter than discs around single stars, reducing their ability to cool quickly enough for GI to operate. Although \citet{Duchene2010} suggested that truncated discs may become more gravitationally unstable and that perturbations from the companion could help trigger collapse, other numerical studies find the opposite trend: dynamical perturbations from the secondary can actually suppress gravitational instabilities \citep[see][and references therein]{Mayer2010}. Last, the metallicity of $\gamma$-Cephei is super-solar, of [Fe/H]$\sim$+0.20 dex, aspect typically linked to the prevalence of the core-accretion scenario for the formation of giant planets \citep{Santos2004}.  Thus, taken together, we believe these considerations make GI an unlikely formation pathway for $\gamma$-Cephei-like planets.

It is worth to emphasise that in any case $\gamma$-Cephei systems are rare: giant S-type planets with $\abin<100$ au have an estimated occurrence rate of 4\% \citep{Hirsch21}, and they represent less than 2\% of the sample of S-type planets in the \citet{Thebault_catlog-2025} Catalog. Thus, although understanding their formation poses an interesting theoretical challenge, the main aim of global models and population synthesis is to reproduced observed demographic trends (which we fairly do, as we will present in Paper III), and not specific systems.

\subsection{Model limitations}
One limitation of this work is the absence of a model for the potential disc photoevaporation driven by the secondary star. This effect could be relevant in close binary systems, where irradiation from the secondary may still affect the outer regions of the primary's disc. As noted by \citet{Rosotti18}, accurately capturing this phenomenon would require full 3D simulations. Nevertheless, we can attempt a rough estimate of the EUV photoevaporation rate by adapting the EUV mass-loss rate prescription of \citet{Matsuyama2003} for external photoevaporation from nearby O/B stars. In their Eq. 16, the EUV mass-loss rate depends on the disc outer edge, on the ionization rate of the external star and on
the distance to the external star. They considered
the distance as that between the dominant star in the Trapezium cluster, $\theta^1$ Ori C, and the proplyds in it, which is of the order of 0.03 pc, and the ionization rate of $\theta^1$ Ori C, which is $\sim10^{49}$ s$^{-1}$. In our context, we can obtain a rough estimate by using the binary separation as the relevant distance, the ionization rate usually considered for sun-like stars, which is $\sim 10^{41}$ s$^{-1}$, and the disc edge as the disc truncation radius. This gives a very low $\dot{M}^{\text{EUV}}_{\text{d}}\sim1\times10^{-10} M_\odot/\text{yr}$, which would affect only insignificantly the most external part of the disc.

Concerning 3D additional effects, we note that out simple 1D disc model cannot capture assymetric effects such as asymetric irraditation or episodic outburst from the companion. This 100 yr-timescale effect might re-set the initial conditions for very close binaries \citep{Poblete2025}. Another hydrodynamic effect which we cannot capture is the appearance of spiral arms \citep{Kley2008}, which might act as local dust traps for planetesimal formation. Such effect deserves further investigation within the context of S-type planet formation.

\begin{figure}
    \centering
    \includegraphics[width=1\linewidth]{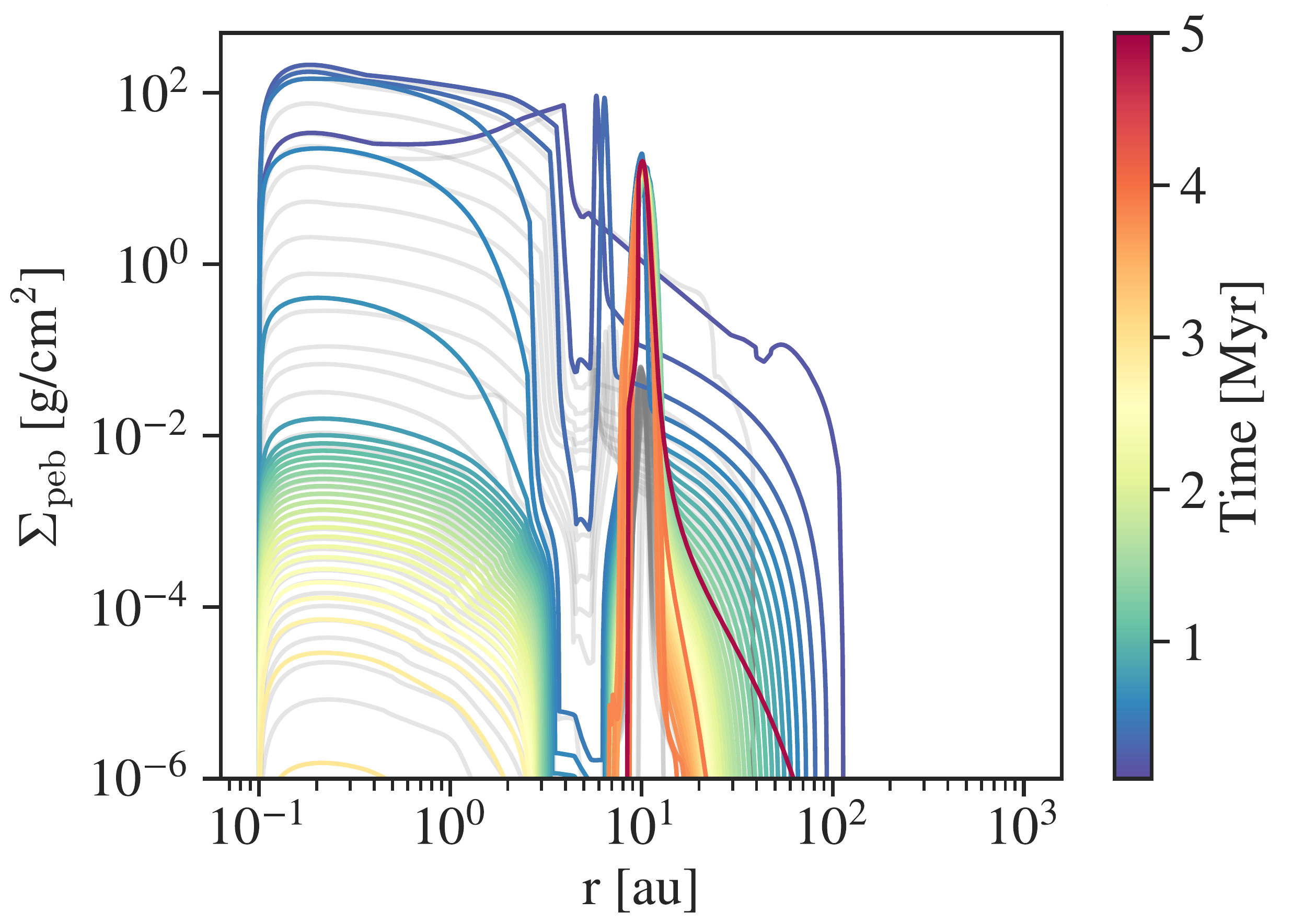}
    \vspace{-0.4cm}
    \caption{Same as Figure \ref{fig_pebble_disc} for the $\abin=100$ au case (and single-star in the grey background), but taking into account the growth of the giant planet at 5 au and the feedback of it on the disc. The evolution of the planet mass is shown in Fig.\ref{fig_growth}, both for the binary and single-star cases.}
    \label{fig:sigmaPeb_w-Planet}
\end{figure}

Regarding the model set-up, for the sake of isolating the effect of the disc truncation on the planet growth by pebble accretion, we considered in this study only in-situ growth of a single embryo. The effect of orbital migration is analysed in Paper II and it is also the default set-up for the population synthesis of the upcoming Paper III. Here we can simply mention that inwards migration just moves the protoplanet to a region where the supply of pebbles lasts longer. This is in principle beneficial for the core growth, but in reality if the protoplanet moves inside the iceline, then the pebble accretion rate drops due to the decrease of the pebbles' size.
Another interesting effect to discuss is how the growth of one planet influences the growth of subsequent planets. In the simulations we presented in Sect.\ref{sec_planet-growth}, we note that for the cases where a giant planet forms, the gap created in the disc produces a density maximum further out, that acts as a pebble-collector \citep[e.g.][]{Guilera17,Guilera20}. We show this in Fig.\ref{fig:sigmaPeb_w-Planet}, which is the same as the one presented on the top right panel of Fig.\ref{fig_pebble_disc}, but considering in this case the growing planet at r=5 au (and its feedback on the disc structure). We note that the pebbles accumulate in the density maximum at r$\approx10$ au. This could be a way to retain more pebbles in the disc and trigger sub-sequent, farther-out, planet formation \citep[as already suggested for single stars by][]{Lau24}. Giant planets should have outer planet siblings, also in binaries.   

\section{Conclusions} \label{sec_conc}
In this paper we introduced the \pairs (Planet formation Around bInaRy Stars) project, which aims to develop the first global formation model for planets orbiting binary stars suited for planet population synthesis. In this first paper, we presented the adaptation of the circumstellar disc to the presence of an external stellar companion, to simulate the formation of S-type planets by pebble accretion. We included the physical effects of tidal disc truncation and heating, and the direct irradiation from the stellar companion into the evolution of a circumstellar disc which undergoes viscous accretion, and internal and external photoevaporation.
The results of this study do not include the gravitational perturbation from the stellar companion on the planetary bodies, which is presented in the accompanying Paper II (Nigioni et al.). 

In this work, we provide an \href{https://github.com/Aryy98/Circumstellar_disc_truncation_radius}{open source script} to compute the outer truncation radius of an S-type disc for any binary parameter (see Sect.\ref{sec_disc_truncation}), based on the fits from \citet{Manara19} to the analytical results of \citet{Arty94}. The disc truncation radius depends on the binary semi-major axis, binary eccentricity, binary mass ratio, and assumed disc viscosity. Both the circumprimary and circumsecondary disc truncation radius can be computed with the provided tool.

We additionally studied how a planetary embryo grows in-situ in these truncated discs, by the accretion of pebbles and gas. We analysed the dependence of the planet growth on different binary parameters, spanning binary separations sampled log-uniformly from 10 to 1000 au, and binary mass ratios from 0.1 to 1. We found that S-type planets attaining a mass larger than Mars form in the whole range of the adopted binary separations, but we note that planet formation gets steadily suppressed for $\abin<160$ au. (Fig.\ref{fig:grid_histo}). The main cause of this detrimental effect on planet formation is the cut-off of the pebble supply from the outer disc due to disc truncation, which starves the growth of the cores (Sects.\ref{sec_dust_disc} and \ref{sec_planet-growth}). In terms of disc truncation, planets more massive than Mars form for $\Rtrunc \gtrsim 3$~au, while the lower limit for planets with $\Mp>10 \,\Mearth$ is $\Rtrunc\approx7$ au (Fig.\ref{fig_param_study1}).
We also find that S-type planets typically form close to their central star in relation to the binary separation and to the disc truncation radius: 75\% of planets more massive than Mars form at $a_{p,0}/\abin$<4\% and $a_{p,0}/\Rtrunc<22\%$. No planet forms for $a_{p,0}/\abin> 30$\% or $a_{p,0}/\Rtrunc > 85\%$. This is also a consequence of the disc tidal truncation: the closer the planet is to the disc truncation radius, the sooner the supply from pebbles stops, halting core growth.

Overall, we showed for the first time that pebble accretion allows for the formation of S-type planets, and we quantified the detrimental effect on planet growth for close binaries when considering the effect of tidal disc truncation. We analyse the added effect of the gravitational perturbation from the stellar companion in the accompanying Paper II, and we will present the results of the first S-type planet population synthesis in Paper III.



\begin{acknowledgements}
    We thank the referee for a very constructive and insightful report that helped us to improve the presentation of the results and discussions. We thank C. Mordasini, A. Kessler, J. Weder and R. Burn for valuable discussions about the Bern Model. J.V. and A.N. acknowledge support from the Swiss National Science Foundation (SNSF) under grant PZ00P2\_208945. MPR is partially supported by PIP-2971 from CONICET
    (Argentina) and by PICT 2020-03316 from Agencia I+D+i (Argentina).
    This work has been carried out within the framework of the NCCR PlanetS supported by the Swiss National Science Foundation. The computations were performed at University of Geneva on the Yggdrasil cluster. This research has made use of NASA's Astrophysics Data System. 
    \textit{Software.} For this publication the following software packages have been used: 
\href{https://matplotlib.org/}{Python-matplotlib} by \citet{Hunter_2007}, \href{https://seaborn.pydata.org/}{Python-seaborn} by \citet{waskom2020seaborn},
\href{https://numpy.org/}{Python-numpy},
\href{https://pandas.pydata.org/}{Python-pandas}
\end{acknowledgements}

\bibliographystyle{aa} 
\bibliography{lit}

\begin{appendix}
\section{The \textit{Bern Model} of planet formation: additional physical assumptions}\label{App_extra-model}

\subsection{Stellar evolution model for the host}

Compared to \citet{Emsenhuber21}, we model the evolution of the host star using the stellar models of \citet{Spada2013} combined with the stellar rotation evolution models of \citet{Johnstone2021}. Following \citet{Venuti2017}, in our simulations we set the central star’s rotation period to 5 days.
It is worth to mention that the stellar evolution grid of \citet{Spada2013} spans from 0.10 $\Msun$ to 1.25 $\Msun$, in steps of 0.05 $\Msun$, so it is also suited for planet formation around the secondary in the mentioned mass range.

\subsection{Calculation of planet properties}

The internal structure of the envelope, during both formation and evolution, is computed by solving the 1D structure equations. In the attached phase, the planetary envelope is in equilibrium with the gaseous disc, so the planets do not have a well-defined radius. Gas accretion is calculated by solving the 1D structure equations of 
\cite{Mordasini12b,Venturini16}, with outer boundary conditions given by \citet{Lissauer09} and maximum gas accretion rate as in \cite{Bodenheimer2013}. For the interior structure of the core, precomputed tables are used instead. More details can be found in \citet{Emsenhuber21}.

When the gas accretion rate exceeds the maximum supply from the disc, the planet transitions to the detached phase \citep{Bodenheimer2000}, and its radius is computed following \citet{Mordasini12b,Mordasini12c}. Since the planetary envelope is no longer connected to the disc, the pressure boundary conditions are adjusted accordingly. These conditions are modified again when the disc disperses, marking the transition to the evolutionary phase \citep{Emsenhuber21}.

The opacity of the planetary envelope, $\kappa$, is computed considering a contribution from the grains, for which we take the interstellar medium grain opacity of \citet{BL94} scaled by a factor of 0.01 to account for grain growth \citep{Mord14, Movshovitz08} as in \citep{Alibert19}; and a contribution from the gas, taken from \citet{Freedman14}. During the evolutionary phase, the envelope is assumed as condensate-free, so only the \citet{Freedman14} opacity tables are considered, for solar composition. Luminosity is calculated using the approach of \citet{Mordasini12b}, which accounts for contributions from accretion, contraction, radioactive decay, and bloating of close-in planets. Further details on these calculations are provided in \citet{Emsenhuber21}. However, in our simulations, we do not account for the bloating of close-in planets.

Once the protoplanetary disc has dissipated, planets enter their evolutionary phase. In close orbits around their host star, they experience strong atmospheric escape due to XUV stellar irradiation. We model this process following \citet{Jin14}, where atmospheric evaporation is initially driven by X-ray irradiation and later by EUV irradiation.

\subsection{Tidal evolution after disc dissipation}

During the evolutionary phase, to account for changes in a planet’s orbital distance, we follow the approach of \cite{BolmontMathis2016}, summarized in the work of \cite{Rao2018}, which is a more refined prescription compared to the standard Bern model presented in \citet{Emsenhuber21}. Specifically, we consider orbital migration due to mass loss (the first term of their eq. 9, excluding stellar mass loss) and the effects of dynamical and equilibrium tides (the last term of their eq. 9). For equilibrium and dynamical tides, we use their eqs. (1) and (5), respectively, and compute stellar rotation using the \citet{Johnstone2021} models, as mentioned earlier.

It is important to note that this model tracks only the evolution of the planet’s semi-major axis, i.e., $\dot{a}/a$, and does not account for changes in other orbital parameters such as eccentricity and inclination. Additionally, each planet is treated independently, meaning planet-planet interactions are not considered during the long-term orbital evolution of multi-planet systems (i.e., from 20 Myr until 10 Gyr, when the N-body integrator is kept running).

\section{Disc model validation}\label{App_model-validation}

To validate our disc model for the time evolution of gas and the vertical structure in circumprimary discs affected by a stellar companion (see Section \ref{sec_methods}), we compare our results with those obtained using PlanetaLP-B \citep{Ronco21}. This 1D+1D code computes the evolution of gaseous discs under various configurations, including circumprimary, circumsecondary, and circumbinary discs, as well as circumbinary discs in hierarchical triple-star systems.

While we calculate the vertical structure at each step of the evolution following the approach of \citet{Nakamoto94}, \citet{Ronco21} computes the vertical structure of the disc following the classical methodology as in \citet{PapaloizouTerquem1999}, with a numerical approach similar to that presented in \citet{Alibert05} and \citet{Migaszewski2015}, but considering an extra tidal heating term in
the energy equation due to the effects of the external companion. This tidal term is analogous to the $Q_{\rm tidal}$ term included in ec. \ref{mid_plane_temp}. We refer the readers to section 2 in \citet{Ronco21} for details.

For this comparison, we compute with PlanetaLP-B the circumprimary disc evolution of the nominal case with binary separation of $a_{\text{bin}}=20$~au (left column of figure \ref{fig_gas_disc}), and considering the same initial gas disc profile and same binary and disc parameters (see Table \ref{tab_gas-disc_No-planets}). Unlike in \citet{Ronco21} where disc truncation is implemented through the inclusion of the tidal torque term in the diffusion equation, we here follow the approach of \citet{Rosotti18,Zagaria2021} and adopt a zero-flux boundary condition at the truncation radius to enable a more direct comparison. Additionally, we consider the same internal and external photoevaporation rates as those considered in this work.

\begin{figure}[h!]
    \centering
    \includegraphics[width=\linewidth]{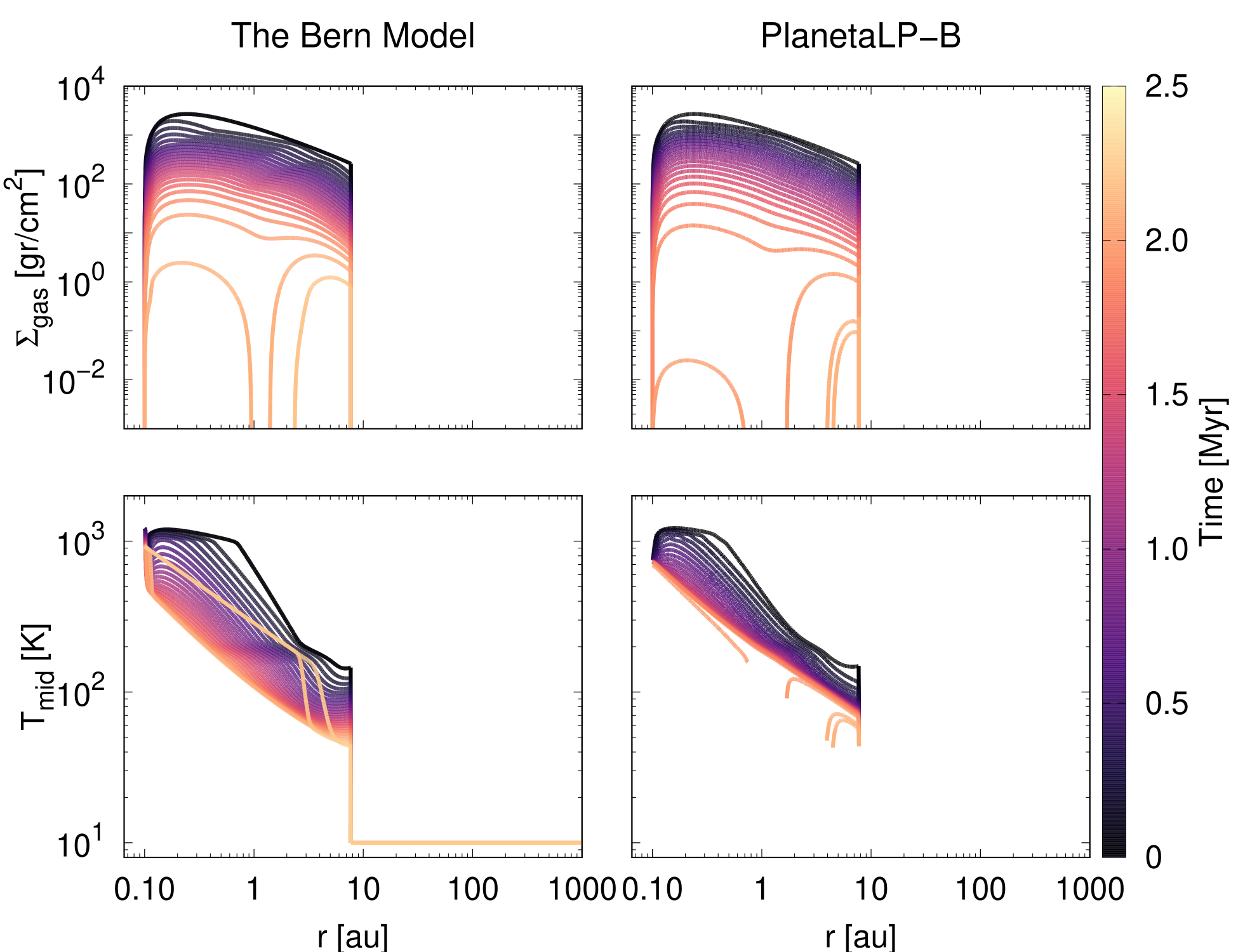} 
        \caption{Comparison of the circumprimary disc evolution for the nominal case with $a_{\text{bin}}=20$ au, between the results obtained in this work and those using the PlanetaLP-B code \citep{Ronco21}. As in figure \ref{fig_gas_disc}, the top panels display the evolution of the gas surface density, while the bottom panels show the evolution of the disc midplane temperature. }
    \label{fig_comp_PlanetaLP}
\end{figure}

Figure \ref{fig_comp_PlanetaLP} shows the comparison between the results obtained with the Bern Model and with PlanetaLP-B (left and right columns, respectively). In general, the results are very similar. Despite some differences, primarily in the disc temperature profiles due to the different numerical approaches, the evolution of the gas surface density is almost identical in both cases, and the disc dissipates within the same timescale. It is worth noting that both models show a temperature increase close to the truncation radius which, as previously mention, is due to the effect of the tidal heating from the stellar companion. The differences in the profiles of the midplane temperature close to the disc dissipation time stem from the different treatment of the vertical structure between the two codes. The Bern Model includes a direct irradiation term (see Sect.\ref{sec_gas-disc}) that ensures a continuos temperature even when the midplane gap is carved, also during the evolution (where the profiles switch to the planetary equilibrium temperature, last two profiles on the left bottom panel of Fig.\ref{fig_comp_PlanetaLP}). 

\section{Disc lifetimes: effect of external photoevaporation} \label{App_externalFoto}
When we analysed the evolution of the S-type gaseous discs in Sect.\ref{sec_gas-disc}, we noted that the disc lifetimes reached a maximum at $\abin \approx 50$ au for our nominal set-up (see Table \ref{tab_gas-disc_No-planets}). In this Appendix we demonstrate that this counter-intuitive effect of having disc lifetimes which do not always decrease with smaller binary separation, is an effect of the external photoevaporation. Indeed, Table \ref{tab_App-lifetimes} displays the disc lifetimes, with and without external photoevaporation, for a range of binary separations. We note that if external photoevaporation is neglected, the disc lifetimes always decrease with decreasing $\abin$, as expected. 

\begin{figure}
    \centering
    \includegraphics[width=\linewidth]{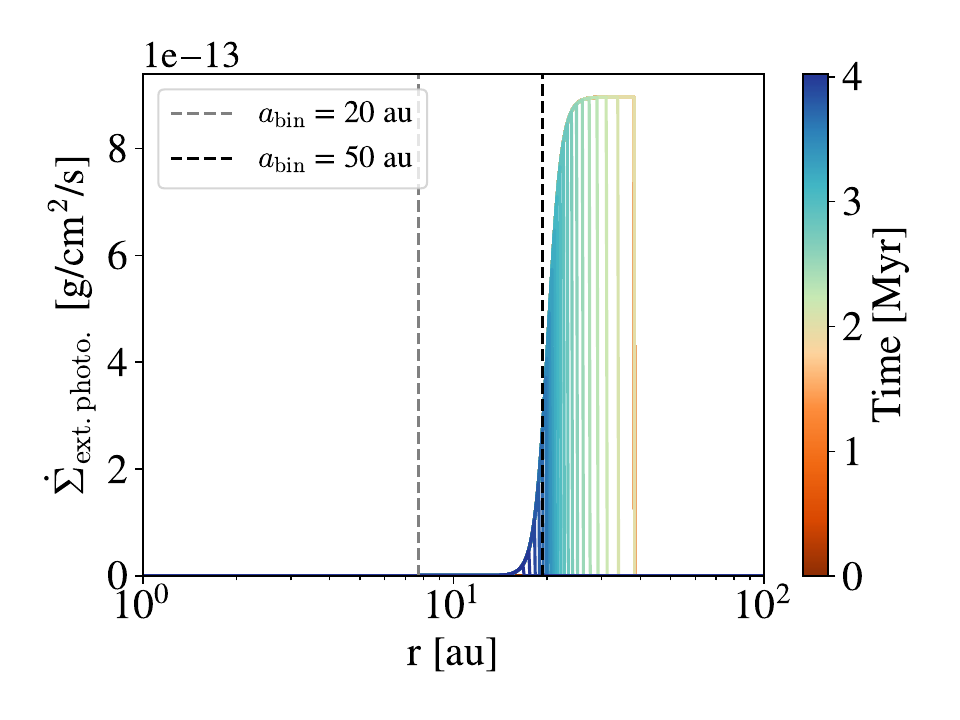}
    \vspace{-0.8cm}
    \caption{External photoevaporation rate as a function of time for the model of \citet{Matsuyama2003}, applied to the nominal disc with $\abin=100$ au (disc truncated at $\Rtrunc=38.5$ au). The 2 extra vertical lines indicate the truncation radius corresponding to a disc with $\abin=20$ (grey) and $\abin=50$ au (black). External photoevaporation is negligible for $r<18$ au.}
    \label{fig:external_photo_rate}
\end{figure}

The external photoevaporation considered in this work, which stems from the Bern Model from \citet{Emsenhuber21} is that of \citet{Matsuyama2003}. The gas removal rate from that simple model is only effective for radial distances $r> 18$ au, as Fig.\ref{fig:external_photo_rate} shows. This is why we see no differences in the disc lifetimes between the model with and without external photoevaporation for truncation radius $\Rtrunc<18$ au (or $\abin<50$ au) in Table \ref{tab_App-lifetimes}. For discs which are truncated at larger radius than 18 au, external photoevaporation becomes increasingly important in removing gas for larger orbital distances (the more distant the gas is from the primary star, the less bound and easier to remove). This is why disc lifetimes get slightly reduced from $\abin$ from 50 au outwards in the nominal setup with external photoevaporation.

\begin{table}[h!]
    \centering
  \begin{tabular}{|c c | c c|}
    \hline
     & &  Nominal & No External \\
     & &   & Photoevap. \\
    \hline
     & & & \\
      $\abin$ [au] & $\Rtrunc$ [au] & $\tau_{\rm disc}$[Myr] & $\tau'_{\rm disc}$[Myr]\\
       & & & \\
       \hline
       20  & 7.7 & 2.1 & 2.1 \\
       30 & 11.6 & 3.5 & 3.5 \\
       40 & 15.4 & 5.1 & 5.1 \\
       50 & 19.3 & 6.3 & 6.7 \\
       60 & 23.1 & 5.5 & 8.3 \\
       70 & 26.9 & 4.9 & 9.9 \\
       80 & 30.8 & 4.6 & 12 \\
       100 & 38.5 & 4.3 & 15 \\
       300 & 116 & 3.9 & 34 \\
       single-star & --  & 3.9 & 34 \\
       \hline
    \end{tabular}
    \vspace{0.2cm}
    \caption{Disc lifetimes for different binary separations for $\ebin=0$, for the nominal model (Table \ref{tab:initialconditions}, middle column) including external photoevaporation ($\tau_{\rm disc}$) and \textit{neglecting} external photoevaporation ($\tau'_{\rm disc}$).}
    \label{tab_App-lifetimes}
\end{table}

\end{appendix}

\end{document}